\documentclass[runningheads]{llncs}
\usepackage[T1]{fontenc}

\usepackage{etoolbox}\usepackage[normalem]{ulem}
\patchcmd{\paragraph}{\itshape}{\textbf}{}{}

\newcommand{\FULLPAPER}{0}
\usepackage[textsize=scriptsize,disable]{todonotes}

\newcommand{\ivan}[1]{\todo[color=orange!30]{\textbf{Ivan: }\\#1}}
\newcommand{\marco}[1]{\todo[color=green!30]{\textbf{Marco: }\\#1}}

\newcommand{\pierp}[1]{\todo[color=red!30]{\textbf{Pierpaolo: }\\#1}}

\usepackage{ifthen}
\usepackage{comment}
\usepackage{tabularx} 
\usepackage{hyperref} 

\ifthenelse{\equal{\FULLPAPER}{0}}{
    \hypersetup{colorlinks = true, allcolors = blue}
    
    \urlstyle{rm}
}{
    \hypersetup{colorlinks = true, allcolors = orange}
}

\usepackage[inline]{enumitem} 
\usepackage[title,toc,titletoc,header]{appendix} 
\usepackage{xspace} 
\usepackage{amssymb}
\usepackage{amsmath}

\ifthenelse{\equal{\FULLPAPER}{0}}{}
{
    \usepackage{amsthm}
    \newtheorem{definition}{Definition}
    \newtheorem{theorem}{Theorem}
}

\usepackage{cryptocode} 
\usepackage{footmisc}

\usepackage[ruled,linesnumbered]{algorithm2e}
\SetKwComment{Comment}{/* }{ */}
\SetKwInput{Witness}{Witness}
\SetKwInput{Instance}{Instance}
\SetKwInput{Parameters}{Parameters}

\makeatletter
\newcommand\footnoteref[1]{\protected@xdef\@thefnmark{\ref{#1}}\@footnotemark}
\makeatother

\usepackage{mathrsfs}

\usepackage{subcaption}

\newcommand{\zksnark}{zk-SNARK\xspace}

\definecolor{spcolor}{gray}{0.28} 
\ifthenelse{\equal{\FULLPAPER}{0}}{
    
}{
    
}

\newcommand{\oursubparagraph}[1]{\bigbreak\noindent\spstyle{{\upshape #1}}}

\newcommand{\repo}{https://github.com/pierpaolodm/ContentPrivacyNFT}

\newcommand{\bc}{\arbiter}
\newcommand{\wallet}{\mathfrak{U}}
\newcommand{\token}{\mathsf{tk}}
\newcommand{\tokens}[1]{\mathsf{tk}[#1]}
\newcommand{\tokenset}{\mathcal{T}}
\newcommand{\pred}{\mathsf{owner}}

\newcommand{\tx}{\mathsf{tx}_\bc}
\newcommand{\store}{\mathsf{store}_\bc}
\newcommand{\intk}{\vartriangleright}

\newcommand{\asset}{\mathsf{a}}

\newcommand{\assetprev}{\mathsf{\asset_{{p}}}}
\newcommand{\assetkey}{\mathsf{{k}_{\asset}}}
\newcommand{\assetsign}{\sigma_{\asset}}
\newcommand{\commassetkey}{\mathsf{c_{\assetkey}}}

\newcommand{\assetenc}{\mathsf{\asset_{{enc}}}}
\newcommand{\fsub}{f_\mathsf{sub}}

\newcommand{\encassetkey}{\mathsf{e_\assetkey}}

\newcommand{\feprot}{\Pi_{\mathsf{FE}}}
\newcommand{\arbiter}{\mathsf{T}}
\newcommand{\seller}{\mathsf{A}}
\newcommand{\buyer}{\mathsf{B}}
\newcommand{\advsubroutine}{\mathsf{defineVerAlg}_{\seller, \buyer}}

\newcommand{\ourfe}{Advertisement-Based Fair Exchange\xspace}
\newcommand{\ourfeprot}{\Pi_{\mathsf{AFE}}}
\newcommand{\Courfeprot}{\Omega_{\mathsf{AFE}}}

\newcommand{\trustparam}{\mathsf{prm}}
\newcommand{\priv}{\asset_{\mathsf{prv}}}
\newcommand{\pub}{\asset_{\mathsf{pub}}}

\newcommand{\ourfesetup}{\mathsf{Setup}}

\newcommand{\ourfeadv}{\mathsf{Advertise}}

\newcommand{\complaint}{\mathsf{Complaint}}

\newcommand{\randomfromset}{{\leftarrow}\vcenter{\hbox{\fontsize{6.8}{7}\selectfont\rmfamily\upshape\$}}~}
\newcommand{\secp}{\lambda}
\newcommand{\negl}{\mathsf{negl}}
\newcommand{\poly}{\mathsf{poly}}
\newcommand{\defeq}{\vcentcolon=}
\newcommand{\boolset}{\{0,1\}}
\newcommand{\pk}{\mathsf{pk}}
\newcommand{\sk}{\mathsf{sk}}
\newcommand{\aux}{\mathsf{aux}}

\newcommand{\trapdoor}{\mathsf{td}}
\newcommand{\G}{\mathbb{G}}
\newcommand{\Z}{\mathbb{Z}}
\newcommand{\cind}{\approx_c}

\newcommand{\N}{\mathbb{N}}
\newcommand{\RO}[1]{\mathcal{H}_{#1}}
\newcommand{\ppt}{\mathsf{PPT}}
\newcommand{\adv}{\mathcal{A}}
\newcommand{\oracle}{\mathcal{O}}
\newcommand{\Ext}{\mathtt{Ext}}
\newcommand{\Sim}{\mathtt{Sim}}
\newcommand{\dist}{\mathcal{D}}
\newcommand{\PRC}[2]{\operatorname{Pr}\left[\begin{array}{c}#2\end{array}\middle\vert \begin{array}{c}#1\end{array}\right]}

    \newcommand{\commscheme}{\Pi_\mathsf{{Com}}}
    \newcommand{\commkey}{\mathsf{ck}}
    
    \newcommand{\commkeygen}{\mathsf{KeyGen}}
    \newcommand{\commcommit}{\mathsf{Commit}}
    \newcommand{\commopen}{\mathsf{Open}}

    \newcommand{\signscheme}{\Pi_\mathsf{{Sign}}}

    \newcommand{\signkeygen}{\mathsf{KeyGen}}
    \newcommand{\signsign}{\mathsf{Sign}}
    \newcommand{\signverify}{\mathsf{Verify}}
    
    \newcommand{\encscheme}{\Pi_\mathsf{{Enc}}}

    \newcommand{\enckeygen}{\mathsf{KeyGen}}
    \newcommand{\encenc}{\mathsf{Enc}}
    \newcommand{\encdec}{\mathsf{Dec}}

    \newcommand{\rel}{\mathcal{R}}
    \newcommand{\stat}{x}
    \newcommand{\wit}{w}
    \newcommand{\crs}{\mathsf{crs}}

    \newcommand{\zksnarkscheme}{\Pi_\mathsf{{ZK\text{-}snark}}}

    \newcommand{\zksnarksetup}{\mathsf{KeyGen}}
    \newcommand{\zksnarkprove}{\mathsf{Prove}}
    \newcommand{\zksnarkverify}{\mathsf{Verify}}

\usepackage[a4paper, margin=1in]{geometry}

\renewcommand{\oursubparagraph}[1]{\smallskip\noindent{{\upshape #1}}}

\newcommand{\CR}[1]{{#1}}
\newcommand{\removeCR}[1]{}

\usepackage{atbegshi,picture}
\AtBeginShipoutFirst{%
  \put(-2.8cm,1.3cm){
    \makebox[\paperwidth][c]{
      \parbox{\dimexpr\paperwidth-2cm}{
        An extended abstract of this work appears in the proceedings of the 
        \emph{24th International Conference on Cryptology and Network Security} (\textsc{CANS’25}). 
        This is the full version. \\A preliminary version of this work appeared with the title 
        ``Public Channel-Based Fair Exchange Protocols with Advertising''.

      }%
    }%
  }%
}

\usepackage{bookmark}
\begin{document}

\title{Decentralized Fair Exchange with Advertising}

    \author{
        Pierpaolo Della Monica \and
        Ivan Visconti \and
        Andrea Vitaletti \and
        Marco Zecchini
    }
    \institute{
        Sapienza University of Rome, Rome, Italy \\
        \email{\{dellamonica,visconti,vitaletti,zecchini\}@diag.uniroma1.it}
    }

    \authorrunning{Della Monica, Visconti, Vitaletti and Zecchini}

\maketitle  

\begin{abstract}
    \pierp{All submissions must be written in English and span no more than 20 pages in Springer’s Lecture Notes in Computer Science (LNCS) format (a Springer template can be found in Overleaf), including title, abstract, and bibliography. The introduction should summarize the contributions of the paper at a level understandable by a non-expert reader and explain the context to related work.

Submitted papers may contain supplementary material in the form of well-marked appendices or a separate file archive (particularly for source code, data files, etc.). Note that supplementary material will not be included in the proceedings. }
Before a fair exchange takes place, there is
typically an advertisement phase with the goal
of increasing the appeal of possessing a digital asset while keeping it sufficiently hidden. 
Advertisement phases are implicit in mainstream definitions, and therefore are not explicitly integrated within fair-exchange protocols. 

In this work we give an explicit definition for such a fair exchange in a setting where parties communicate via broadcast messages only (i.e., no point-to-point connection between seller and buyer is needed). Next, we construct a fair-exchange protocol satisfying our new definition using {\zksnark}s and relying on mainstream decentralized platforms (i.e., a blockchain with smart contracts like Ethereum and a decentralized storage system like IPFS).

Experimental results confirm the practical relevance of our decentralized approach, paving the road towards building decentralized marketplaces where users can, even anonymously, and without direct off-chain communications, effectively advertise and exchange their digital assets as part of a system of enhanced NFTs.
\pierp{Dovremmo nell' abstract forse porre più l'accento su ``public channel-baed'' in accordo al cambio di titolo.}
\ivan{abstract da cambiare visto che (come discutemmo mesi fa), non è la nostra definizone a introdurre per la prima volta l'advertisement}

\end{abstract}

\section{Introduction}
\label{sec:introduction}
An exchange protocol involves two parties, Alice and Bob, each interested in exchanging her/his asset with that of the other party.
A fair exchange protocol guarantees that, at the end, either the assets are exchanged or none of them is transferred. 
If Alice and Bob are interested in exchanging a {\em digital} asset through a protocol run over a communication network, without the involvement of a trusted third party, a fair exchange is impossible. This is due to a general impossibility result proven by Cleve~\cite{STOC:Cleve86} on fairness in two-party protocols.

\paragraph{Advertising before exchanging: the  case of physical assets.}
In the physical world, advertising an asset typically consists of granting temporary access to it (e.g., before an auction, the bidders can watch a picture at a specific distance and can read documents proving possession of it). In general, such advertisement phases do not compromise the value of the asset because physical assets cannot be easily cloned or exploited (e.g., the bidders that have temporary access to the auctioned asset cannot easily reproduce an identical copy).

\paragraph{Fair exchange in the NFT Era.}
Blockchains can realize a basic decentralized trusted party, and this suffices to bypass Cleve's impossibility. Indeed, smart contracts for Non-Fungible Token (NFT) allow to fairly exchange the ownership of an asset for tokens. 
However, digital assets can not be advertised by just showing their contents, since they can be trivially copied.  Indeed, a general criticism about NFTs for artworks is that regulating the possession of publicly available digital assets is often considered of modest interest, since the asset can also be enjoyed without owning it. 
Systems for managing NFTs guarantee that there is one owner of an asset, but still anyone can download and enjoy to some extent the asset. 
In general, it is useful to provide mechanisms allowing a party, simply represented by a public key (i.e., there is no information about who is behind the public key and thus no idea about his reputation/reliability), to  (appealingly and reliably) advertise her desire to exchange her asset giving some specific guarantees to buyers while, at the same time, keeping the asset sufficiently hidden (e.g., an artist producing high-value digital pictures shows only an appealing short preview of the picture, but then there is no guarantee that she has the full asset). 

It is therefore essential when considering the fair exchange of digital assets, to model the possibility of keeping the original asset private while publicly advertising only a corresponding restricted preview (that is, a ``low'' quality version of the asset) that is appealing enough to stimulate the interest of others.

\paragraph{TTP-based (or reputation-based) approaches.}
An obvious way to realize the above goal of having advertisable fair exchange protocols, consists of introducing a Trusted Third Party (TTP) 
both in the advertisement phase and during the actual exchange. 
In this scenario, users and/or companies rely on a well-known and reputed TTP to advertise and trade their digital assets. 
Such a TTP manages the assets and implements access control policies that limit the exposed information about the assets to be traded. At the same time, the TTP advertises the assets through previews and performs the exchanges, commonly for money. 
There are two types of trust assumptions: (a) the TTP is trusted to correctly advertise the asset (e.g., the preview is correctly computed)  and (b) the exchange will be fairly carried out (e.g., if the buyer transfers the amount specified by the seller, then the buyer will receive the asset and the seller will receive the money). 

In the wild, nowadays, there exists a vast variety of marketplaces selling assets (i.e., services that offer Alice to advertise her asset and then to fairly sell Alice's asset for Bob's money) all sticking with the above centralized architecture. 
For the case of physical assets, notable examples can be found in the context of car dealerships, where the party willing to buy a car can get temporary access to it (e.g., performing a test drive before buying a car) or in real estate, where the party can get temporary access to a building.
Also for the case of digital assets, there are renowned examples of marketplace allowing to advertise and fairly exchange: (a) high-resolution images (e.g.,  {Getty Images}, 
{Shutter Stock}) advertised by an image preview (e.g., blurring and/or watermarking are applied to the image), (b) streaming video (e.g., {Netflix}, {Amazon Prime}) advertised by a video trailer or (c) streaming audio (e.g., {Spotify}, {Audible}) where just a portion of the audio is made publicly available for the advertisement. 

The above trust towards the TTP is not necessarily a leap of faith. Indeed, in real-world scenarios, if Alice or Bob detects a misbehavior from the TTP then some evidence can be provided damaging the reputation of the TTP, inducting other users to change service, or even pursue lawsuits. Essentially, the TTP is just a company that runs a business that, as usual, is  economy driven and a good reputation is one of the key factors for its success.

\paragraph{Risks of relying on services based on reputation. }
While reliable servers (i.e., TTPs or companies with good reputations) are the main choice today to advertise and trade digital assets, there are several important  limitations that make the state of affairs quite unsatisfying. 
Servers could be successfully attacked becoming unreliable against their will (e.g., images deposited on a server could be stolen in case of a data breach).
Attacks are likely to happen when servers manage assets of great value. Moreover, in specific contexts server can be censored by a government, limiting in turn the free exchange of assets.
A server (if any) that is considered resilient to the above vulnerabilities is likely to be expensive.

In turn, users might not be satisfied by the policies of the server (e.g., an artist selling her songs might be unhappy with the high fees of the service and the poor quality of the advertisement) or by the limited security and availability. 

\subsection{Fair Exchange in the Decentralization Era}
For the reasons discussed above, it is natural to consider a setting in which a user wishes to deploy her own service and act as the server herself. However, establishing a trustworthy reputation in such a decentralized setting requires both time and effort. During the initial bootstrapping phase, buyers may be reluctant to engage in transactions with a seller who lacks any form of endorsement or reputation from a TTP.
This motivates the need to design systems that avoid dependence on centralized or costly intermediaries, while still ensuring both a reliable advertisement phase and a fair exchange of digital assets.

To frame the discussion, consider an example where a buyer, Bob, wishes to purchase a digital asset (e.g., an image, video, or piece of music) from a seller, Alice. Bob wants to ensure that he receives the desired asset in exchange for his payment, while Alice wants to be confident that he will receive the correct payment in return. A fair exchange protocol should provide the following informal guarantees:
    
    \noindent(\textbf{I}) The buyer (Bob) must be guaranteed that if he participates in the protocol but does not receive the agreed-upon digital asset, then he retains full control of his funds (i.e., no coins are lost or transferred).
    
    \noindent(\textbf{II})  The seller (Alice) must be guaranteed that if she does not receive the correct payment, then the buyer learns nothing about the asset beyond what was intentionally made public during the advertisement phase (e.g., the preview). No sensitive or valuable information about the asset should be leaked.
    
    \noindent(\textbf{III})  The advertising of the digital asset should occur in a public and unique manner. That is, the seller should be able to publish the preview once, after which it remains publicly accessible. This reflects the natural structure of real-world marketplaces, where advertisements are broadcast to all potential buyers rather than being shared privately on a per-request basis. This property has several implications: 
    \begin{enumerate*}[label=(\textbf{\alph*})]
        \item it ensures that the seller is not vulnerable to  DoS attacks from malicious buyers repeatedly requesting previews;
        \item it avoids requiring the buyer to be constantly online to respond to preview requests;
        \item it prevents malicious sellers from impersonating honest ones by copying their publicly available previews and falsely claiming ownership of the asset.
    \end{enumerate*}
    Thus, to achieve transparency and resilience, the entire interaction, including both the advertisement and the exchange, should be conducted without a point-to-point communication between buyer and seller. 

\label{sec:feitw}
\label{sec:related_work}
\paragraph{Other works in the literature.}
Starting with the popular solution based on Hashed TimeLock Contracts~\cite{HTLC}, 
fair exchange via blockchains has been widely explored in the scientific literature~\cite{Maxwell15,ESORICS:BanDziMal16,CCS:CGGN17,CCS:DziEckFau18,ASIACCS:EckFauSch20}
\cite{FileBounty,ESORICS:HLTWW21,CCS:LYHMGZ21,CCS:TSZMKB24,smartZKCP,cryptoeprint:2025/388,cryptoeprint:2025/059}. All such works essentially focus on an asymmetric setting considering a seller named Alice who would like to sell a digital asset for an amount of cryptocurrency provided by a buyer Bob, giving this type of fair exchange the name of \emph{contingent payment}. 
Note that the impossibility result of~\cite{STOC:Cleve86} is circumvented since Alice and Bob are now equipped with a trusted (decentralized) infrastructure.
Those results do not communicate the asset $\asset$
upfront (thus protecting it) and guarantee to a specific buyer that the asset $\asset$ is a witness of an NP relation that will be received when the payment is complete.

These protocols typically address a ``witness-selling'' problem, where a buyer offers an amount of coins for a witness value $w$ of an NP relation (e.g., knowledge of the discrete logarithm), such that $\rel(x,w) = 1$ for a shared instance $x$ between the seller and the buyer.
Therefore, it is natural to consider the NP relation as a formalism to capture interesting properties of the witness (i.e., the asset $\asset$) to be exchanged, thereby using such properties to ``advertise'' the asset $\asset$ to the buyer.

For the sake of clarity, we sketch here the common pattern used in the above works. Let us suppose that Alice aims to sell a digital asset $\asset$. 
At first, Alice and Bob somehow agree on the asset $\asset$ to exchange (e.g., Alice publicly claims that she owns the asset $\asset$ and Bob asks for it).
Then (1) Alice, without disclosing the asset $\asset$ to exchange, sends to Bob a preview $\assetprev$ of the asset $\asset$ also stating the price for it. 
Then (2), Bob, after having looked at the preview $\assetprev$ and agreed on the price, asks Alice to begin a fair exchange protocol.
After that, (3) Alice computes the encryption $\assetenc$ of $\asset$ with a symmetric key $\assetkey$ (i.e., $\assetenc = Enc(\assetkey,\asset)$), the commitment $\commassetkey$ of the symmetric key (i.e., $\commassetkey = Com(\assetkey)$), and (4) privately sends $\assetenc$ and $\commassetkey$ along with a consistency proof for the following NP relation:
\[
\begin{array}{ccc}
    \rel(x \defeq (\commassetkey,\assetenc,f,\assetprev),w \defeq (\assetkey,\asset)) = 1 \,\,
    \text{if and only if}\\
    \left(\commassetkey = Com(\assetkey) \land \assetenc = Enc(\assetkey,\asset) \land \assetprev = f(\asset) \right)     
\end{array}
\]
where $w$ is the witness for the relation $\rel$, given $x$ as the instance. 
If Bob verified this consistency proof correctly, then (5) Bob deployed a smart contract that will transfer a certain amount of cryptocurrency to the user capable of opening the commitment (i.e., $\commassetkey$) received by Alice. Finally, (6) Alice opens $\commassetkey$ and redeems the cryptocurrency by publicly disclosing $\assetkey$, thereby also allowing Bob to obtain it. Bob
can now retrieve $\asset$ (i.e., Bob can obtain $\asset = Dec(\assetkey,\assetenc)$).

\paragraph{Security and usability issues of standard fair exchanges.} 
According to the above discussion on the blueprint that all fair exchanges share,  the seller transmits $\assetenc$, $\commassetkey$ and a consistency proof over a point-to-point channel that would limit the usability of contingent payments, especially in the presence of large amounts of fair exchanges done by a party (e.g., requiring parties to be simultaneously online and/or to coordinate how they exchange messages directly and privately). 

In the non-optimistic approach, as in~\cite{CCS:CGGN17} the seller, after having sent privately to a single buyer, in the above step (4), the commitment and the encryption,  has also to compute and send an ad-hoc consistency proof to the buyer to provide guarantees on the witness (i.e., the asset). This can seriously affect the seller's performance, especially in the presence of high demand (i.e., many buyers' requests) and/or when the underlying NP statement being proved is complex (e.g., due to a large statement size or the inherent complexity of the NP relation). 

In the optimistic approach such as the one described in~\cite{CCS:DziEckFau18}, 
the consistency proof is missing and the unsatisfied buyer, after having verified that the received asset does not match the expected properties, must compute and send a proof-of-misbehavior. According to~\cite{ASIACCS:EckFauSch20}, this can be an expensive task for the buyer and the smart contract, therefore leading to high costs. 
Furthermore, in both approaches, the buyer may interact with \emph{fake sellers} who do not possess the desired asset, thereby resulting in wasted effort and time. Such vulnerabilities have also been informally observed and discussed in~\cite{smartZKCP}.

A very recent work~\cite{CCS:VanSonThy24} also considers the setting of fair exchange in the context of advertising. Their approach employs adaptor signatures as the underlying cryptographic primitive to properly formalize the advertisements phase. In their protocol, the buyer creates a presignature on both the claim and the associated blockchain transaction. The seller can convert this presignature into a valid signature, and thereby obtain payment, only if she possesses a valid witness for the claim (i.e., the asset being sold).
Although this approach provides a specific notion of fairness also with  advertisements, it relies on the assumption of point-to-point  communication between the parties. As in other prior works, also the work of~\cite{CCS:VanSonThy24} suffers from the ``{fake seller}'' problem, wherein multiple malicious sellers may initiate the protocol with no intention of completing it. Due to the private nature of the interaction, there is no public evidence of such aborts, preventing others from being aware or responding to this misbehavior.

In practice, existing fair exchange protocols are never executed exclusively through broadcast messages sent to decentralized platforms, and thus they require parties to communicate directly to each other (e.g., via point-to-point communication). As a result, property~\textbf{(III)} is consistently violated: interactions often rely on private communication, and the advertisement phase is neither modeled as a separate component nor implemented in a one-time, publicly verifiable way. 
Moreover, current definitions typically omit an explicit treatment of the advertisement phase and fail to provide formal security guarantees for it. Consequently, property~\textbf{(II)} is not explicitly addressed. Fully addressing these gaps remains an open problem.



\subsection{Our Results}
We show how to realize fair exchange protocols for a decentralized marketplace where parties publicly, efficiently and cheaply advertise and exchange their assets without off-chain interactions with other parties, hence securely working through decentralized platforms.

We give a definition that explicitly combines a fair-exchange protocol with a prior advertising phase, a direction that has been recently and independently explored also by~\cite{CCS:TSZMKB24}.
In our definition, the malicious buyer that does not transfer the required tokens does not learn anything about an asset except what has been on purpose leaked by the seller through the desired advertised information. At the same time, the definition guarantees that malicious sellers will not obtain tokens from the honest buyer unless the buyer gets the entire asset that is legitimately corresponding to the advertised version. 

In contrast to prior work, our definition explicitly models the advertisement phase as a distinct and one-time step, and mandates that all interactions through broadcast messages sent to decentralized platforms. This addresses directly the shortcomings discussed above.

Next, we present an optimistic construction that combines a \zksnark (off-chain, but still publicly verifiable as it is uploaded on IPFS) as a consistency proof for advertising the asset, along with an optimistic fair exchange over a blockchain, applied in the context of the NFT markets. As discussed above, the NFT market for digital artworks suffers from a fundamental limitation: the traded assets are publicly accessible, even to those who do not own the corresponding NFT. This significantly reduces the appeal of such applications.
This is because the security of the smart contract relies on the transparency of the underlying blockchain, allowing any user to autonomously verify the system's status by accessing all its data.
Improving the management of NFTs for confidential assets, potentially enables a new wave of interest in such an application context. Thanks to our construction, an NFT that manages confidential assets permits only the owner of the NFT to have access to the asset managed by such NFT, while interested buyers can access only a preview, with a guarantee that the preview is correctly computed and this limited access can encourage buyers to purchase the asset. If interested, buyers can then purchase the NFT and obtain access to the asset.

Our construction works as follows.
\begin{enumerate}[label=(\textbf{\alph*})]
    \item As advertisement, the seller publishes on a decentralized storage platform the \emph{preview of the asset},  an \emph{encryption of the asset}, a \emph{commitment on the encryption key} used to encrypt the asset and a \emph{\zksnark} proving that she knows both the encryption key and the asset corresponding to the preview. 
    
    \item The seller then generates a pair of Diffie–Hellman (DH) \emph{ephemeral keys} and deploys a \emph{smart contract} to facilitate the fair exchange. In the contract, the seller publishes the \emph{sale price} of the asset (i.e., the amount of cryptocurrency she wishes to receive), a commitment to the encryption key, and a DH message (i.e., her public contribution to the DH key exchange), while keeping the corresponding secret key private.
    
    \item Once the \zksnark is verified, and it is confirmed that the commitment to the encryption key stored in the smart contract's internal state matches the one used in the \zksnark, a user (acting as the buyer) proceeds as follows: she generates a DH message along with the corresponding secret key, initiates the fair exchange protocol by locking the sale price (i.e., the specified amount of cryptocurrency) in the smart contract, and publishes her DH message.
    
    \item The seller satisfied by the locked cryptocurrency, first computes a DH \emph{session key} and computes a \emph{cyphertext} (with this freshly computed session key) of a message that consists of the above encryption key (i.e., the committed one) concatenated with the randomness used to compute the commitment. Finally, she publishes the resulting ciphertext in the smart contract.
    \item The buyer, computing the  DH session key, will decrypt the ciphertext uploaded by the seller obtaining the encryption key and the randomness used in the commitment. 
    
    If the encryption key matches the one committed in the decentralized storage, the buyer will correctly decrypt the asset and afterward the locked cryptocurrency will be automatically unlocked on the blockchain and the seller will receive the payment.  Otherwise, the buyer can initiate a dispute by publishing in the smart contract his secret corresponding to his DH contribution. 
    
    The smart contract performs the following steps: \begin{enumerate*}[label=(\textbf{\arabic*})]
        \item it verifies the correctness of the secret uploaded by the buyer by recomputing the buyer's DH message and checking that this corresponds to the one previously updated by the buyer;
        \item it computes the DH session key combining the DH message of the seller with the secret of the buyer;
        \item it uses the session key to get through decryption of the very same encryption key computed by the buyer along with the randomness for the commitment, and, finally, 
        \item it verifies whether this corresponds to the commitment stored in its internal state.
    \end{enumerate*} 
    
    If the computed commitment does not correspond to the one that was previously updated by the seller, it means that the seller is trying to obtain the buyer's cryptocurrencies without disclosing the correct encryption key. In this case, the smart contract can penalize the seller (for example, by burning the NFT) to discourage bad behaviors. If the computed key corresponds to the committed key, then the buyer is cheating and the smart contract can discard the request or, similarly to the seller's case, penalize the buyer. 
\end{enumerate}
   
Note that by assigning a percentage of future revenues to
previous owners, they can be economically incentivized not to disclose the encryption keys.

We have benchmarked our construction, focusing our experiments on the management of NFTs that represent images available in renowned collections (e.g., BAYC). The goal of our experiments is to evaluate the performance of the proposed construction for NFTs of confidential artworks, to show the viability of our solution and the advantages of the new definition.
We have conducted two main experiments, that consist of a) proving the ability to compute the proofs to advertise the confidential images and b) evaluating the costs of the smart contract extending the ERC-721 standard and implementing the fair exchange. Both experiments confirm the viability of the proposed approach, and a comparison of layer-1 (Ethereum) and layer-2 (Optimism)  deployment and operation costs for the smart contract is shown to allow the selection of the most appropriate technology according to the value given to the asset.

\section{Preliminaries}
    Due to space limitations, standard notation of cryptographic primitives are deferred to Sec.~\ref{app:preliminaries}.
\subsection{Tokens} 
\label{sec:tokens}
Taking inspiration from \cite{C:BenKum14},
we define \emph{tokens}  with the following properties:
\begin{enumerate*}[label=({\arabic*})]
\item a token owner is the only user in possession of it, and it is ensured that no other entity can simultaneously own the same token;
\item tokens can be freely transferred from a sender to a receiver (i.e., the sender is no longer the token owner while the receiver becomes the new token owner) and the received token's validity can be immediately verified and confirmed;
\end{enumerate*}
finally, we assume that each user should have access to a \emph{wallet}  $\wallet$ to accumulate \emph{tokens}. 

More formally, we denote with $\tokenset$ the set of all possible tokens with the same value.
We denote with $\tokens{n} \defeq \{\token_1,\token_2,\ldots,\token_n\}$ a set of $n$ tokens such that $\token_i \in \tokenset$ for $1 \leq i \leq n$.
We highlight that the comparison between two sets of tokens (e.g., $\tokens{n} > \tokens{m}$) is a comparison based on their total quantity.
Specifically, $\tokens{n} > \tokens{m}$ holds if and only if $n > m$, with $n,m \in \N$.

We explicitly write $\tokens{n} \intk \wallet$ to indicate that the tokens $\{\token_i\}_{1 \leq i \leq n}$ are owned by the user who controls the wallet $\wallet$ (i.e., the user knows the secret key associated with $\wallet$, enabling them to authorize transactions that spend these tokens). Following standard blockchain assumptions, ownership is defined by the ability to spend, which in turn requires knowledge of the corresponding secret key. As is common in prior work, we abstract away the cryptographic details of key management and refer to token ownership in terms of wallet access.
We denote by $\pred$ a predicate that takes as input a set of tokens $\tokens{n}$ and a wallet $\wallet$, and outputs a bit in $ \{0,1\}$. The output is $1$ only if  all tokens in $\tokens{n}$ are contained in $\wallet$ (i.e., $ \forall i \in \{1, \ldots, n\},\ \token_i \in \wallet$); otherwise, it outputs $0$.

\subsection{Blockchain}
\label{sec:blockchain}
Introduced in 2008, 
 a blockchain is a distributed ledger with growing lists of records (i.e., blocks) securely linked together via cryptographic hashes~\cite{EC:GarKiaLeo15}.

The status of the system evolves through transactions. 
A transaction can transfer digital currency from a sender user to a receiver user. Every user that interacts with the system has a pair of keys of $\signscheme$  generated with the $\signscheme.\signkeygen$ algorithm. 
According to Sec.~\ref{sec:tokens}, we provide every user with a \emph{wallet} containing tokens and the blockchain maintains a ledger of all wallets.
A user $\mathsf{U}$, that has $(\pk_\mathsf{U}, \sk_\mathsf{U})$, signs every transaction that she wants to add to the ledger, with $\sk_\mathsf{U}$. We denote the wallet of the user $\mathsf{U}$ as $\wallet_\mathsf{U}$.
A transaction is added to the ledger if the signature is correctly verified using the public key $\pk_\mathsf{U}$ and if $\mathsf{U}$ has enough tokens in $\wallet_\mathsf{U}$ to perform that operation.

\paragraph{Global clock.} 
In addition to managing transactions and maintaining a decentralized ledger, the blockchain can also serve as a global clock. Loosely speaking, we define time in the system as a value maintained by the blockchain in the form of a global clock, observable by all participants. This clock advances as time progresses (i.e., according to the creation of new blocks), ensuring that all parties share a consistent view of the current time, denoted by $t$. 
More formally, since we consider blockchains modeled as a synchronous model of communication, we implicitly assume that the blockchain acts also as a global clock~\cite{TCC:KMTZ13}, with protocols executed in rounds and every party aware of the current round. Thus, parties can expect messages to be received at a certain time. This assumption in the blockchain domain 
is indeed widely employed in other works in this area~\cite{SP:ThyMal21,SP:TaiMorMaf21,SP:ThyMalMor22,CCS:GMMMTT22}.

\paragraph{Smart contracts.}
In 2015, Ethereum 
 introduced the concept of smart contracts, i.e. it allows users to deploy arbitrary programs that not only define the conditions for spending money, but also enable the deployment of generic programs that use the blockchain to reach consensus on a program state. Therefore, user transactions can be viewed as inputs to these \emph{smart contracts}. In summary, Ethereum has introduced and realized the concept of a virtual machine on top of a blockchain.

Taking inspiration from~\cite{FCW:ADMM14,SP:ThyMal21} and leveraging the capabilities of smart contracts, we equip the blockchain with two interfaces through which a (sender) user can interact to execute transactions: 
\begin{enumerate*}[label=(\arabic*)]
    \item $\mathsf{tx}(\mathsf{U},\tokens{n}, t')$: given as input a (receiver) user $\mathsf{U}$ (where $\mathsf{U}$ represents the user identified by the public key $\pk_\mathsf{U}$), a set of $n$  tokens $\tokens{n}$, and (optionally) a future time instant $t'$ such that $t' \geq t$ (where $t$ denotes the current time), 
    this transaction locks the tokens $\tokens{n}$  until time $t'$ on the blockchain (i.e., on a smart contract). After $t'$, the tokens are transferred to the (receiver) user $\mathsf{U}$. 
    This transaction models the behavior of a simple \emph{time-lock transaction}. Note that if $t'= \bot$ (i.e.,  $t'$ is not specified), this denotes a simple transaction over the blockchain to transfer $\tokens{n}$ tokens to the user $\mathsf{U}$.

    \item $\mathsf{store}(m)$: given a message $m \in \boolset^*$ as input, the message $m$ is made visible to all parties, similar to posting it on a public bulletin board.
\end{enumerate*}

\section{\ourfe}
\label{sec:definitions}
\removeCR{
    Due to space limitations, standard notation of cryptographic primitives are deferred to Sec.~\ref{app:preliminaries}.
}

Here, we define \ourfe.
Following the approach of other works on fair exchange discussed in Sec.~\ref{sec:related_work}, we focus on the asymmetric case, where one party exchanges an asset for money. We consider this scenario as the most interesting case, while treating the symmetric case, in which both parties exchange assets, as an obvious extension.

In accordance with the definition of fair exchange {in Sec.~\ref{sec:fair_ex}}, the protocol involves three actors. The first two are Alice and Bob, who want to exchange their respective assets. 
In the protocol, Alice will be the seller who has advertised her asset, and Bob will be the buyer who wants to pay a certain amount of money to obtain this asset.
Finally, the third actor is the TTP referred to as $\arbiter$\footnote{Jumping ahead, we adopt for our construction a restricted TTP $\arbiter$ publicly maintaining all the information and assuming that $\arbiter$ performs its operations correctly.}.


For $\ppt$ algorithms $\mathsf{A}$, $\mathsf{B}$ and $\mathsf{C}$,
we denote by $[x,y,z] \leftarrow [\mathsf{A}(a), \mathsf{B}(b), \mathsf{C}(c)](d)$ the random process obtained by having $\mathsf{A}$,$\mathsf{B}$ and $\mathsf{C}$ interact on
common input $d$ 
and on (private) auxiliary inputs $a$, $b$ and $c$, respectively,
and with independent random coin tosses for $\mathsf{A}$, $\mathsf{B}$ and $\mathsf{C}$. 
The values $x$, $y$ and $z$ represent the outputs of $\mathsf{A}$, $\mathsf{B}$ and $\mathsf{C}$, respectively, after the interaction. 

We define an asset and an asset preview as bit strings, namely $\asset, \assetprev \in \boolset^*$, and a preview function $f: \boolset^* \rightarrow \boolset^*$ as an efficient deterministic function that, given in input an asset $\asset$, outputs $\assetprev $. 


\begin{definition}\label{def:ourfe}
    An \ourfe  $\ourfeprot$ for a preview function $f$ 
    is a three-party protocol run by a $\ppt$ seller $\seller$, a $\ppt$ buyer $\buyer$  and a TTP $\arbiter$
    satisfying \emph{Completeness}, \emph{Seller Fairness}, \emph{Buyer Fairness} and \emph{Broadcast Communication only} and that works as follows:
    \begin{itemize}
        \item $\trustparam \leftarrow \ourfesetup(1^\secp)$ is a randomized algorithm that takes as input the security parameter $\secp$, and outputs a string $\trustparam \in \boolset^*$ that represents  (possibly trusted) parameters.
        
        \item $(\priv,\pub) \leftarrow \ourfeadv(\trustparam, \asset, \token_{\mathsf{thr}})$ is a randomized algorithm that takes as input $\trustparam$, an asset  $\asset \in \boolset^*$ and $\token_{\mathsf{thr}} \in \N^+$ and outputs a pair $(\priv,\pub)$.

        \item $[\alpha, \beta] \leftarrow [\seller(\priv), \buyer(\tokens{n}), \arbiter](\trustparam,\pub)$ is a three-party protocol.  $\seller$ and $\buyer$ exchange $\priv$ and $\tokens{n}$, respectively, where $\tokens{n}$ are the buyer's tokens used to purchase $\priv$. 
        $\buyer$ engages the protocol with $\seller$ after evaluating $\trustparam$ and $\pub$.

        At the end of the protocol, $\seller$ outputs $\alpha$ and $\buyer$ outputs $\beta$. The value $\bot$ as output indicates that the party has rejected the execution. 
        
    \end{itemize}

\oursubparagraph{Completeness.} 
 Given an honest seller $\seller$,  with wallet $\wallet_\seller$, and an honest buyer $\buyer$, with wallet $\wallet_\buyer$ such that $\tokens{n} \intk \wallet_\buyer$\footnote{\label{ftn:tkinwallet} We recall that, according to Sec.~\ref{sec:tokens}, the notation $\tokens{n} \intk \wallet_\buyer$ indicates that the tokens $\tokens{n} = \{\token_i\}_{1 \leq i \leq n}$ are owned by the user who has access to the wallet $\wallet_\buyer$ (i.e., the user who knows the secret key to transfer the token inside the wallet), in this case, $\buyer$. Namely, that $\tokens{n}$ are contained within $\buyer$'s wallet.}, for any asset $\asset \in \boolset^*$ and any $\token_{\mathsf{thr}} \in \N^+$, the following probability is equal to $1$:
\[
\PRC{
    \trustparam \leftarrow \ourfesetup(1^\lambda)\\
    (\priv,\pub) \leftarrow \ourfeadv(\trustparam, \asset, \token_{\mathsf{thr}})\\
    }{
   [\tokens{n}, \priv] \leftarrow [\seller(\priv), \buyer(\tokens{n}), \arbiter](\trustparam,\pub) \\
   \land \pred(\tokens{n}, \wallet_{\seller})
}
\]
We recall that, we denote with $\pred$ a predicate that, taking as input any set of tokens $\tokens{n}$ and any wallet $\wallet_\mathsf{U}$ accessible by a user $\mathsf{U}$, it outputs $1$ if all the tokens are in the wallet $\wallet_\mathsf{U}$ (i.e., $ \forall i \in \{1, \ldots, n\}, \token_i \in \wallet_\mathsf{U}$), otherwise  $0$
.

\oursubparagraph{Seller Fairness.}
Given an honest seller $\seller$, for any asset $\asset \in \boolset^*$ and any $\token_{\mathsf{thr}} \in \N^+$, for all $\ppt$ malicious buyer $\buyer^*$ with a wallet $\wallet_\buyer$ and $\tokens{n} \intk \wallet_\buyer$\footnoteref{ftn:tkinwallet}, there exists a PPT algorithm $\Sim_{\buyer^*}^\arbiter$  with oracle access to $\arbiter$ such that, for any $\ppt$ distinguisher $\dist$, the following probability is at most $ \negl(\secp)$:
\[ 
\begin{array}{ll}
    & \left| 
    \PRC{
    \trustparam \leftarrow \ourfesetup(1^\lambda)\\
    (\priv,\pub) \leftarrow \ourfeadv(\trustparam, \asset, \token_{\mathsf{thr}})\\
    
    [\bot, \beta] \leftarrow [\seller(\priv), \buyer^*(\tokens{n}), \arbiter](\trustparam,\pub)
    }{
    \dist(\asset,\beta) = 1
    } \right.
    \\\\
    & - \left. \PRC{
     \beta \leftarrow \Sim_{\buyer^*}^\arbiter(f(\asset), \token_{\mathsf{thr}})
    }{
    \dist(\asset,\beta) = 1
    }\right|
\end{array}
\]
Namely, this property requires that, at the end of the protocol, regardless of the output of $\buyer^*$ (denoted generically as $\beta$ in the above probability), if the output of $\seller$ is $\bot$ (i.e., $\seller$ aborts the execution), the probability that $\buyer^*$ learns something more than $f(\asset)$ and $\token_{\mathsf{thr}}$ is negligible. 

\oursubparagraph{Buyer Fairness.}
Given an honest buyer $\buyer$, with a wallet $\wallet_\buyer$ such that $\tokens{n} \intk \wallet_\buyer$\footnoteref{ftn:tkinwallet}, for all $\ppt$ malicious seller $\seller^*$ the following probability is at most $\negl(\secp)$:
\[
    \PRC
    {
        \trustparam \leftarrow \ourfesetup(1^\lambda)\\
        (\priv,\pub) \leftarrow \seller^*(\trustparam)\\
        
        [\alpha, \bot] \leftarrow [\seller^*(\priv), \buyer(\tokens{n}), \arbiter](\trustparam,\pub)
    }
    {
        \exists i \in \{1, \ldots, n\}\text{ s.t. } \token_i \not \in \wallet_\buyer
    }
\]
Namely, this property requires that, at the end of the protocol, regardless of the output of $\seller^*$ (denoted generically as $\alpha$ in the above probability), if the output of $\buyer$ is $\bot$ (i.e., $\buyer$ aborts the execution), the probability that $\buyer$ has lost a token from $\tokens{n}$ in $\wallet_\buyer$ is negligible.

\oursubparagraph{Broadcast Communication only.}
\pierp{check and add comments in the proof}
\marco{added to the proof}
All communication between the seller $\seller$ and the buyer $\buyer$ is conducted exclusively via a broadcast messages towards decentralized platforms. That is, all messages exchanged between the parties must be forwarded via $\arbiter$, who ensures that every message is immediately made publicly visible.

\ivan{per mantenere anche il contributo definizionale del nostro lavoro, secondo me potremmo innestare la proprietà di comunicare solo via un bulletin board; in fondo è questo il punto di forza che ci permette di distinguere poi la nostra costruzione rispetto a quelle precedenti}

\end{definition}
Note that the definition is parametrized for a preview function $f$. The description of this function is included in $\pub$ during the execution of $\ourfeadv$.
 
The fairness properties for both sellers and buyers are inspired by~\cite{CCS:TSZMKB24}. Specifically, in the context of \emph{contingent payment} protocols, there is a natural asymmetry between the seller and the buyer in terms of guarantees provided by the fairness property, as highlighted in~\cite{CCS:TSZMKB24}. In general, the properties of Def.~\ref{def:ourfe} grant the same level of security of the definition in~\cite{CCS:TSZMKB24} but our definition explicitly devises an advertisement operation for the seller, making it more suitable for a decentralized marketplace of confidential assets.  

For seller fairness, the focus is on preserving the confidentiality of the seller's private information (denoted as $\priv$ in our case) when the protocol between the seller and the buyer does not conclude successfully. If the honest seller aborts the protocol, no (non-trivial) information about $\priv$ should be revealed. To formalize this, we adopt a simulation-based definition of fairness, where a malicious buyer cannot distinguish whether she is interacting with a simulator or the actual seller, and the simulator does not know the seller's secret (i.e., $\priv$ in our case).

Following the definition provided in~\cite{CCS:TSZMKB24}, we adopt a relaxed notion of security to ensure buyer fairness (i.e., protect the buyer's tokens). While a stronger definition would require the inclusion of a simulator even in this case, we align with the approach taken in previous works and opt for a more traditional game-based definition. Indeed, 
buyer fairness does not require confidentiality for the buyer's tokens, $\tokens{n}$, and ensures that no token is transferred to the seller at the end of the protocol between the seller and the buyer, when the protocol does not conclude successfully; the entire amount of tokens for the exchange is still owned by the buyer.

%
Note that, this ``asymmetry'' between the seller and the buyer fairness (due to the \emph{contingent payment} scenario)  can also be observed in other cryptographic protocols such as the zero-knowledge and soundness properties of zero-knowledge proofs. 
W.r.t. the buyer fairness of~\cite{CCS:TSZMKB24}, we do not focus on the tokens that the adversary holds at the end of the protocol, and, instead, we capture the property that the balance of the honest party (i.e., the buyer, $\buyer$, in our definition) at the end of the protocol remains unchanged, when the buyer is honest and outputs $\bot$. This variation makes our definition meaningful also in those scenarios where a token can be destroyed. 
Due to space limitations, further considerations on how to easily extend Def.~\ref{def:ourfe} to the ``symmetric case'' (i.e., both parties advertise and then exchange an asset) are deferred to {~Sec.~\ref{app:notes-on-definition}}.

\CR{
Finally, note that the three-party protocol in Def.~\ref{def:ourfe} does not specify the number of messages exchanged between the parties. This generality allows for optimistic constructions, where parties can optionally raise complaints to $\arbiter$ about misbehavior during protocol execution. Such an approach is valuable in practice, enabling efficient execution when all parties behave honestly, while still providing security guarantees through the complaint mechanism when needed.

More formally, our Advertisement-Based Fair Exchange definition accommodates a complaint mechanism within the protocol, allowing a party  to provide cryptographic evidence to $\arbiter$ demonstrating that the other party has deviated from the prescribed protocol. This mechanism can be instantiated by enabling a party to submit specific values (e.g., secret keys or intermediate protocol values) that allow $\arbiter$ to verify the complaint's validity, and punish the (other) party that deviates from the protocol.
The complaint mechanisms usually operate within a temporal framework defined by three critical time bounds. Let $t_{\textsf{init}}$ denote the protocol initiation time (when assets are committed or tokens locked), $t_{\textsf{comp}}$ denote the complaint deadline, and $t_{\textsf{final}}$ denote the protocol conclusion time (when assets or tokens are irreversibly transferred). These time points must satisfy $t_{\textsf{init}} < t_{\textsf{comp}} < t_{\textsf{final}}$ to ensure that: (1) parties have sufficient time to execute the protocol, (2) complaints can be filed before irreversible transfers occur, and (3) $\arbiter$ has adequate time to evaluate complaints and take appropriate action.

We emphasize that Def.~\ref{def:ourfe} naturally accommodates such complaint mechanisms without requiring explicit modification, as the fairness properties already capture the necessary security guarantees. The definition's flexibility allows various implementations of complaint mechanisms, e.g. including both single-round and multi-round complaint procedures.
}



\section{Construction}
\label{sec:constructions}
We present a construction of an Advertisement-Based Fair Exchange, in which the TTP is implemented through a smart contract involving the fair exchange of confidential and authenticated Non-Fungible Tokens (NFTs).
Fair exchanges in a public marketplace can involve a wide variety of digital items, ranging, for example, from videos and software vulnerabilities to trained neural networks and images. 
Our proposed construction specifically targets the fair exchange of NFTs managing confidential and authenticated assets on a blockchain. 
Enabling the trading of such tokens addresses a significant issue within the blockchain domain, as already deeply discussed in Sec.~\ref{sec:introduction}. 
In the following, we show how an Advertisement-Based Fair Exchange is perfectly suited to manage the advertisement and the trading of authenticated and confidential assets (i.e., realizing de facto a market of NFT managing confidential assets) by requiring only a blockchain equipped with smart contracts (e.g., Ethereum
).

\subsection{Advertisement-Based Fair Exchange for NFTs}
\label{sec:constructionNFT}
W.l.o.g., we simplify the representation of an NFT omitting some values such as the association of an owner publicly visible on the blockchain or the unique ID assigned to every NFT (e.g., as in Ethereum ERC721~\cite{bauer2022erc}). Instead, we primarily focus on the asset itself. The management of the other properties associated with the NFT can be easily integrated into the construction.

\paragraph{Parties and asset description.}
According to Def.~\ref{def:ourfe}, we consider an Advertisement\allowbreak-Based Fair Exchange where a seller $\seller$ owns an NFT managing a (confidential) asset $\asset^\star$ that she wants to publicly advertise and then trade with an interested buyer $\buyer$ for some tokens $\tokens{n}$. The NFT and the Advertisement-Based Fair Exchange involve a TTP $\arbiter$, which is a blockchain\footnote{\label{ftn:sctobc}We generically refer to $\arbiter$ as a blockchain rather than a smart contract. This mainly because, depending on the specific blockchain and how NFTs are managed, an NFT might be controlled by multiple smart contracts and/or with a different logic.} (see Sec.~\ref{sec:blockchain}). Note that, according to Sec.~\ref{sec:blockchain}, $\arbiter$ works correctly.
We focus solely on the management of the asset $\asset^\star$ and do not consider other boundary properties that characterize the NFT. This (1) simplifies the treatment and (2) enables us to embrace different types of NFT that can be defined and managed differently according to different blockchains (e.g., Ethereum ERC721 
or Algorand ASA
).

We consider the asset $\asset^\star \in \boolset^*$ composed of a set of bits, a public key and a signature that attest the authenticity of this asset. We adopt this type of representation because it seems very general and natural for a wide variety of assets in the NFT domain.
That is, denoting the asset to exchange with $\asset^\star=(\asset,\pk, \assetsign)$, we consider $\asset \in \boolset^*$ as a generic set of bits, $\pk$ as a public key of $\signscheme$ (see\removeCR{~Sec.~\ref{sec:sign}} the full version) and  $\assetsign$ as a digital signature of $\signscheme$. 
A well-formed asset $\asset^\star$ always has the following property: $\signscheme.\signverify(\pk,\asset,\assetsign) = 1$. 
Note that, $\pk$ represents the public key of the creator (and sometimes also the current owner) of the asset $\asset$ (e.g., a public key of the artist who created the image $\asset$, or the company that issued the ticket $\asset$). 

\paragraph{Remarks on $\arbiter$ when represented by a blockchain.}
The TTP, denoted as $\arbiter$, involved in the exchange between parties, is represented by a blockchain\footnoteref{ftn:sctobc}. 
According to Sec.~\ref{sec:blockchain}, all information held by $\arbiter$ must be considered public. In other words, once $\arbiter$ stores or uses a value for computation, that value is publicly available to everyone (i.e., it is equivalent to posting the value on a public bulletin board). Furthermore, as discussed in Sec.~\ref{sec:blockchain}, we equip $\arbiter$ with two interfaces with which each user can interact, namely $\tx$ and $\store$. 

Finally, according to the definition of an \ourfe (see Def.~\ref{def:ourfe}), oracle access to $\arbiter$ is required. When $\arbiter$ is represented by a blockchain, this requirement naturally follows, as it implies the ability to interact with $\arbiter$ through the $\mathsf{store}$ and $\mathsf{tx}$ interfaces (according to Sec.~\ref{sec:blockchain}), and to query the blockchain and retrieve all operations and information maintained within it. This aligns perfectly with the behavior of most real-world blockchains today.

\paragraph{Description of algorithms.}
 We recall that $\ourfeprot$ is parameterized by a preview function $f$. We denote our construction of $\ourfeprot$ as \hypertarget{Courfeprot}{$\Courfeprot$}.  The construction $\Courfeprot$ is parametrized by a function $f$ that takes as input $\asset^\star \defeq (\asset, \pk, \assetsign)$ and 
 outputs the tuple $(\fsub(\asset), \pk)$ composed of the output of the transformation of $\asset$ (leveraging $\fsub$ that is described in $f$) and the $\pk$, that is basically forwarded from the input.
 
\paragraph{Protocol setup.} In the Advertisement-Based Fair Exchange defined in Sec.~\ref{sec:definitions}, there is a setup algorithm namely $\ourfeprot.\ourfesetup$, this algorithm outputs a string $\trustparam \in \boolset^*$ representing the trusted parameters. $\ourfeprot.\ourfesetup$ is run by an additional trusted party (not the blockchain $\arbiter$) that disappears after the execution of the algorithm.
In our construction, $\trustparam$ is a pair composed of the common reference string  $\crs$ generated with $\zksnarkscheme.\zksnarksetup$, that is built starting from a relation $\rel$ and the commitment key $\commkey$ generated with $\commscheme.\commkeygen(1^\secp)$ (i.e., $\trustparam \defeq (\commkey,\crs)$)\footnote{We introduce this trusted party only for convenience, to generate the trusted parameters used by the protocol. In practice, one can instantiate the \zksnark in the RO model~\cite{EC:BCRSVW19} and derive the commitment key via the RO as well (i.e., by standard hashing into the group technique). This eliminates the need for any trusted setup. Our use of a CRS-based setup is aligned with our experimental evaluation, but the same construction can be easily realized entirely in the RO model, avoiding this other party.}.
The \zksnark, for which the $\crs$ is defined, is based on the following NP relation \hypertarget{main-relation}{$\rel$}:
\[
\begin{array}{lc}
&\rel 
    \left(
        x \defeq (\assetenc, \assetprev 
        ,\commassetkey,\commkey),
        w \defeq (\assetkey,r)
    \right) = 1 
   \; 
\text{if and only if}
\\
&\left(
    \begin{array}{cc}
         (\asset||\assetsign) = \encscheme.\encdec(\assetkey, \assetenc) 
         \land  1 = \signscheme.\signverify(\pk, \asset, \assetsign)  \\ 
         \land  \assetprev = (\fsub(\asset),\pk) = f((\asset, \pk, \assetsign))  
         \land  \commassetkey = \commscheme.\commcommit(\commkey, \assetkey; r)
    \end{array}
\right)
\end{array}
\]

\paragraph{Advertisement of the asset.} 
The definition of $\ourfeprot.\ourfeadv$ in Sec.~\ref{sec:definitions} allows advertising the asset $\asset^\star$ that the seller wants to trade for tokens. 
In particular, given
a trusted parameter $\trustparam$,
an authenticated asset $\asset^\star = (\asset, \pk, \assetsign)$
and an amount of tokens $\token_{\mathsf{thr}} \in \N^+$,
the seller computes a pair of values $(\priv, \pub)$ where $\pub$  is the information to advertise $\asset^\star$ and $\priv$ is the information to trade $\asset^\star$.
That is, the seller:
\begin{enumerate*}[label=(\textbf{\arabic*})]
    \item provides guarantees about the asset $\asset^\star$ through $\pub$, and
    \item  advertises $\asset^\star$ by publishing  $\pub$, that includes also the price, namely the amount of tokens  that the seller wants to exchange for the asset. 
\end{enumerate*}

In Alg.~\ref{alg:adv}, we formally describe how $\ourfeprot.\ourfeadv$ works in our construction, thus, how $(\pub,\priv)$ are computed and how $\asset^\star$ is advertised through $\pub$. 
Moreover,  $\pi$ is generated with a \zksnark
, according to definition of \zksnark given in{~Sec.~\ref{app:preliminaries}}. 
Alg.~\ref{alg:adv} invokes at line 4 an internal sub-routine called $\advsubroutine$ that, taking as input the amount of tokens that the seller wants to exchange for the asset (i.e., $\token_{\mathsf{thr}} \in \N^+$), outputs a verification algorithm $V_\seller$ for the seller  and a verification algorithm $V_\buyer$ for the buyer. The verification algorithm $V_\seller$, taking as input a set of tokens $\tokens{k}$, outputs $1$ if $k$ is equal to $ \token_{\mathsf{thr}}$; otherwise, it outputs $0$. The verification algorithm $V_\buyer$ enables the buyer to verify that she has correctly received the asset $\asset^\star$.

\begin{algorithm}
\caption{The algorithm  $\ourfeprot.\ourfeadv$ for $\asset^\star$.}\label{alg:adv}

\KwIn{Trusted parameters $\trustparam$ such that $\trustparam \defeq (\commkey,\crs)$, an asset $\asset^\star \defeq (\asset, \pk, \assetsign)$ and 
an amount of tokens $\token_{\mathsf{thr}} \in \N^+$ that the seller wants to exchange for the asset.}
$\assetprev \leftarrow  f(\asset^\star);$
$\assetkey \leftarrow \encscheme.\enckeygen(1^\secp);$
$\assetenc \leftarrow \encscheme.\encenc(\assetkey,\asset||\assetsign);$
$r \randomfromset \Z_q$\\
$\commassetkey \leftarrow \commscheme.\commcommit(\commkey,\assetkey; r);$
$\pi \leftarrow \zksnarkscheme.\zksnarkprove(\crs,(\assetenc,\assetprev,\commassetkey,\commkey),(\assetkey,r))$\\
$V_\seller, V_\buyer \leftarrow \advsubroutine(\token_{\mathsf{thr}});$ 
$\pub\defeq((\assetenc,\assetprev,\commassetkey,\commkey),\pi,V_\seller, V_\buyer);$
$\store(\pub)$\\
\KwOut{$(\pub,\priv\defeq \assetkey\|r)$}
\end{algorithm}

\paragraph{Exchange protocol.}
In Fig.~\ref{fig:exprot}, we describe the protocol that the seller $\seller$ and the buyer $\buyer$ engage to exchange their assets (i.e., the NFT $\asset^\star$ and the set of tokens $\tokens{n}$). According to the definition in Sec.~\ref{sec:definitions}, before engaging the protocol, the buyer $\buyer$ evaluates  $\pub\defeq((\assetenc,\assetprev,\commassetkey,\commkey),\pi,V_\seller, V_\buyer)$. In our construction, this evaluation works as follows: 
first, the buyer $\buyer$ evaluates $V_\seller$ from $\pub$ by simply looking at $V_\seller$ and $\assetprev$ and deciding whether it is worth purchasing the asset. Then, $\buyer$ evaluates the correctness of the algorithm $V_\buyer$ from $\pub$.
Finally, regarding  the remaining part of $\pub$ (i.e., $(\assetenc,\assetprev,\commassetkey,\commkey)$ and the proof $\pi$), this involves  $\buyer$ checking whether $\zksnarkscheme.\zksnarkverify(\crs,(\assetenc,\assetprev,\commassetkey,\commkey),\pi) = 1$.

The protocol is detailed in Fig.~\ref{fig:exprot}. 
In a nutshell, the parties first execute an ephemeral Diffie-Hellman key exchange, obtaining the shared key $k_{\seller\buyer}$. Next, the buyer $\buyer$ locks into $\arbiter$ the tokens $\tokens{n}$ for a time $t'$. Following this, the seller $\seller$, first checks if $\tokens{n}$ (locked by $\buyer$) fulfills her requirements (i.e., $V_\seller(\tokens{n}) = 1$), and then using the shared key $k_{\seller\buyer}$, encrypts $\priv$ to securely complete the exchange of the asset $\asset^\star$.
We recall that $\priv$ is composed of two substrings that are:
\begin{enumerate*}[label=(\textbf{\arabic*})]
    \item the (symmetric) key $\assetkey$ used to encrypt $\asset^\star$ (i.e., $\encscheme.\encenc(\assetkey, \asset||\assetsign)$), and
    \item the randomness $r$ used for the commitment of $\assetkey$ (i.e., $\commscheme.\commcommit(\commkey, \assetkey; r)$).
\end{enumerate*}

Then, the seller publishes the encryption $\encassetkey \defeq \encscheme.\encenc(k_{\seller\buyer}, \assetkey)$ in $\arbiter$ by invoking $\store(\encassetkey)$. Finally, $\buyer$ checks the correctness of $\encassetkey$ (i.e., the buyer runs an internal sub-routine $V_\buyer$), and in case of failure she complains with $\arbiter$; otherwise, if $\buyer$ obtains the correct $\assetkey$, she ends the protocol and $\seller$  gets the locked tokens $\tokens{n}$, after $t'$. To complain with $\arbiter$, we equip the blockchain with an additional interface $\complaint_\arbiter(\sk_\buyer)$ that can be invoked by $\buyer$ executing Alg.~\ref{alg:complaint}. \CR{Specifically, Alg.~\ref{alg:complaint} first verifies the complaint's timeline and the buyer's identity using the provided secret key. It then derive the shared key between buyer and seller, decrypt the encrypted asset key, and verify the commitment. If verification fails (indicating seller misbehavior), the arbiter returns tokens to the buyer.} Note that this can be fairly done since we focus on blockchains where it is possible to run arbitrary smart contracts, according to Sec.~\ref{sec:blockchain}. 
\begin{figure}[!htb]
    \begin{center}
    \fbox{\parbox{.83\linewidth}{
\pseudocodeblock[linenumbering]{    \seller(\priv\defeq(\assetkey\|r) ) \< \< \buyer(\tokens{n}) \\
            [.1\baselineskip][\hline]
            \< \< (\pk_\buyer,\sk_\buyer) \leftarrow \mathsf{Gen}(1^\lambda); \store(\pk_\buyer)\\
            \< \< \tx(\seller,\tokens{n}, t')\\
            \textbf{if}~V_\seller(\tokens{n}) = 0: \textbf{output}~\bot\\
            (\pk_\seller,\sk_\seller) \leftarrow \mathsf{Gen}(1^\lambda);
            \store(\pk_\seller) \\
            k_{\seller\buyer} \leftarrow \RO{\ell}(\sk_\seller\cdot\pk_\buyer)\\
            \encassetkey \leftarrow \encscheme.\encenc(k_{\seller\buyer}, \assetkey\|r)\\
            \store(\encassetkey)\\
            \< \< k_{\seller\buyer} \leftarrow \RO{\ell}(\sk_\buyer \cdot \pk_\seller)\\
            \< \< \textbf{if}~V_\buyer(\commassetkey,\encassetkey,k_{\seller\buyer}) = 0: \\
            \< \< \quad\textbf{output}~\complaint_\arbiter(\sk_\buyer)
        }
    }}
    \end{center}
    \caption{The exchange protocol performed by $\seller$ and $\buyer$, namely a construction of the following protocol $[\seller(\priv), \buyer(\tokens{n}), \arbiter](\trustparam,\pub)$ in Def.~\ref{def:ourfe}.}
    \label{fig:exprot}
\end{figure}

\paragraph{On the notation used in Fig.~\ref{fig:exprot}.}
Here, we clarify some aspects related to the protocol described in Fig.~\ref{fig:exprot}.
The function $\RO{\ell}$ is a \emph{random oracle}  where $\ell$ is the random oracle output size of the following dimension $\ell \defeq |\mathcal{K}_e|+|\mathcal{M}_c|$.
Within the protocol,  $\mathsf{Gen}(1^\lambda)$ is used to generate a key pair $(\sk,\pk)$. Specifically, the randomized algorithm $\mathsf{Gen}$, when invoked, first selects  $\sk \randomfromset \Z_q$ and then computes the public key as $\pk = (\sk \cdot G) \in \G$ with $G$ a generator of $\G$ that is a group of prime order $q$. This naturally leads to the Diffie-Hellman key exchange.

According to the discussion in Sec.~\ref{sec:blockchain}, we recall that we consider time as a value maintained by $\arbiter$ in the form of a global clock, observable by all parties,  which increases as time progresses and that we denote the current time with $t$. 

\paragraph{Description of $V_\buyer$.} 
The internal subroutine  $V_\buyer$, executed by the buyer to verify the correctness of the received information $\encassetkey$, consists of the steps:
\begin{enumerate}

    \item It checks whether $\arbiter$ has received $\encassetkey$ (i.e., $\encassetkey = \bot$ in $\arbiter$). If $\arbiter$ has received $\encassetkey$, the algorithm continues to the next step. If not, it checks whether the current time $t$ (i.e., the time maintained by $\arbiter$) is greater than $t_\mathsf{comp}$\footnote{\label{ftn:tcomp}The value $t_\mathsf{comp}$ is an instant of time after the time lock transaction invoked by $\buyer$ (i.e., $\tx(\seller,\tokens{n}, t')$) but shorter than $t'$.}. If $t \leq t_\mathsf{comp}$, the algorithm returns $1$, indicating that $\seller$ still has time to send the encryption $\encassetkey$. If $t > t_\mathsf{comp}$, it returns $0$.
    
    \item It decrypts $\encassetkey$ as $(\assetkey'\|r') \leftarrow \encscheme.\encdec(k_{\seller\buyer}, \encassetkey)$, and returns the result of the committment opening, $\commscheme.\commopen(\commkey,\commassetkey,\assetkey',r')$.
\end{enumerate}

\begin{algorithm}
\caption{This is executed by $\arbiter$ after calling $\complaint_\arbiter$.  $\arbiter$ can reuse all data that were previously stored in it. 
When invoked in the protocol described in Fig.~\ref{fig:exprot}, $\arbiter$ knows $\pub,\pk_\seller,\pk_\buyer$ and $\encassetkey$.
}\label{alg:complaint}

\KwIn{The value $\sk_\buyer$ representing the secret key of $\buyer$.}
\If{$t > t_{\mathsf{comp}}\footnoteref{ftn:tcomp} \And \encassetkey = \bot$}
{   $\arbiter$ gives the tokens  $\tokens{n}$ back to $\buyer;$
    \Return $\bot$
}
$\pk_\buyer' \leftarrow (\sk_\buyer \cdot G) \in \G$\\
\If{$\pk_\buyer' \neq \pk_\buyer$}
\Return
    
$k_{\seller\buyer} \leftarrow \RO{\ell}(\sk_\buyer\cdot\pk_\seller)$; 
$(\assetkey\|r)\leftarrow \encscheme.\encdec(k_{\seller\buyer},\encassetkey)$\\
\If{$\commscheme.\commopen(\commkey,\commassetkey,\assetkey, r) \neq 1$}
{   $\arbiter$ gives the tokens  $\tokens{n}$ back to $\buyer;$
    \Return $\bot$
}
\end{algorithm}

\begin{theorem}
\label{thm:opti-constr}
    Given a deterministic preview function $f$ and  $\arbiter$ instantiated by a smart contract allowing general-purpose computations, consider the following 4 ingredients:
    \begin{enumerate*}[label=\normalfont(\textbf{\arabic*})]
        \item $\encscheme = (\enckeygen, \encenc, \encdec)$ is a semantic secure  encryption scheme,
        \item $\signscheme = (\signkeygen,\allowbreak \signsign, \signverify)$ is an EUF-CMA signature scheme,
        \item $\commscheme = (\commkeygen, \commcommit, \commopen)$ is a commitment scheme and,
        \item $\zksnarkscheme =\allowbreak (\zksnarksetup, \zksnarkprove, \zksnarkverify)$ is a \zksnark 
        for the relation \hyperlink{main-relation}{$\rel$}.
    \end{enumerate*}
    Additionally, assume the Decisional DH assumption holds.
    
    Then, \hyperlink{Courfeprot}{$\Courfeprot$} is a secure $\ourfeprot$ protocol for $f$ in the random oracle model.
\end{theorem}
The proof of Theorem~\ref{thm:opti-constr} is in{~Sec.~\ref{app:theorem-proof}}.

Accordingly, we defer to{~Sec.~\ref{app:notes-on-contruction}}  for more details on  \texorpdfstring{$\Courfeprot$}{the construction}. Specifically, in{~Sec.~\ref{app:notes-on-contruction}}  we look at the amount of information maintained by $\arbiter$, discuss the efficiency of $\Courfeprot$, and consider the implications for previous asset owners.

\CR{\paragraph{One advertisement to rule them all.}
We observe that our construction, although simple, based on the definition in Def.~\ref{def:ourfe} of $\ourfeprot$, effectively overcomes the problems, described before in Sec.~\ref{sec:feitw}, that typically affect both the seller and buyer in state of the art fair exchange protocols when instantiated in the wild. We quickly recall that these problems arise when an advertisement phase is required before the asset exchange and when the seller has to engage with multiple buyers.

More in details, regarding the issue of a DoS attack against the seller, we see that this problem does not arise in our construction $\Courfeprot$. This is because the seller generates the proof, which may be computationally heavy, only once for a given asset, rather than repeatedly for each buyer who wishes to engage in the exchange. Moreover, if the current seller is a previous buyer who has already purchased the asset and now wishes to resell it, she does not need to recompute the proof. Instead, she reuses the proof she received during their initial purchase and proceed directly with the exchange protocol, which remains lightweight. In fact, for an honest seller, the protocol involves only a Diffie-Hellman key exchange followed by an encryption of a secret key. In a nutshell, this greatly reduces the computational workload for the seller, as the potentially heavy proof generation is required only once. Additionally, if the seller is not the first owner of the asset, she can reuse the proof she obtained from the previous sale.

From the buyer’s perspective, we address a potential replay attack scenario, where multiple smart contracts (thus sellers) attempt to trade the same asset, without actually known the asset itself but just by copy and paste $\pub$ (in our construction the \zksnark proof). Since the buyer does not need to privately communicate with the seller and the proof is public, other buyers can inform them about the seller’s reputation. Furthermore, additional protection measures can be easily put into place. It is possible to include in the relation \hyperlink{main-relation}{$\rel$}, which generates the \zksnark proof, a reference to the specific smart contract address where the exchange will take place. In this way, the buyer can verify the address before engaging in the exchange, ensuring they interact with the correct smart contract and preventing them from wasting time on an incorrect/malicious one.}

\section{Experiments}
\label{sec:experiments}
In this section, we demonstrate the feasibility of our approach for advertising and selling NFTs that represent images. This choice is motivated by the natural alignment with the challenges affecting this technology, as discussed in Sec.~\ref{sec:introduction}, and the significant real-world importance of such assets.

We consider images from renowned NFT collections (\url{https://ebutemetaverse.com/nft-image-size}), confirming experimentally the viability of our Advertisement-Based Fair Exchange for NFTs (detailed in Sec.~\ref{sec:constructions}). We focus on two aspects: (a) the evaluation of the \zksnark  setup and the \zksnark proof generation, in Alg.~\ref{alg:adv} line 3, and (b) the costs of the smart contract that implements the fair exchange between the parties.
In our experiment, we use different image sizes that correspond to the dimensions of the images in the NFT collections (e.g., Bored Ape Yacht Club  (BAYC)), ranging  from $128 \times 128$ to  $631 \times 631$ pixels. 

The bilinear resize function,
that halves both the width and height of the original image, is used to generate the preview. 
According to~\cite{SP:DVVZ25} we focus solely on the resize as the preview function. That is, in~\cite{SP:DVVZ25} the authors, also focusing on images, demonstrate that during the computation of a \zksnark when additional operations, such as hashing the image (as in our case), are involved, changing the preview function introduces negligible overhead, namely using other preview functions (e.g., cropping or grayscale) would yield similar complexity.

A resize applied to an original small-size image might be considered of limited practical value. However, other transformations can be applied and we will discuss how to extend our framework to support larger images (up to $\sim$30 MP) considering the work in \cite{EPRINT:DEH24,SP:DCB25,SP:DVVZ25}. The ability to deal with large-size private images will open new opportunities for the NFT market. Current NFT standards, certify the ownership of a NFT, but the corresponding image can be easily accessed and copied. Keeping private the high-quality original image can incentivize the development of a ``new'' NFTs market, where not simply the ownership is guaranteed, but also the private access to the high-quality original image, thus overcoming a relevant issue in the current NFTs market. 

In order to demonstrate the feasibility of our approach, in the following we consider the relatively small image sizes used in real-world NFT collections. 
We conducted our experiments on an Intel(R) Core(TM) i9-10900 CPU @ 2.80GHz processor with 20 cores and 32 GB of RAM\footnote{All the code is available  on~\url{\repo}.}. We deferred the discussion about technical choices for the experiments to{~Sec.~\ref{app:notes-on-experiments}}.

\begin{figure}[htbp]
    \centering
    \begin{subfigure}[b]{0.45\linewidth}
        \centering
        \includegraphics[width=\linewidth]{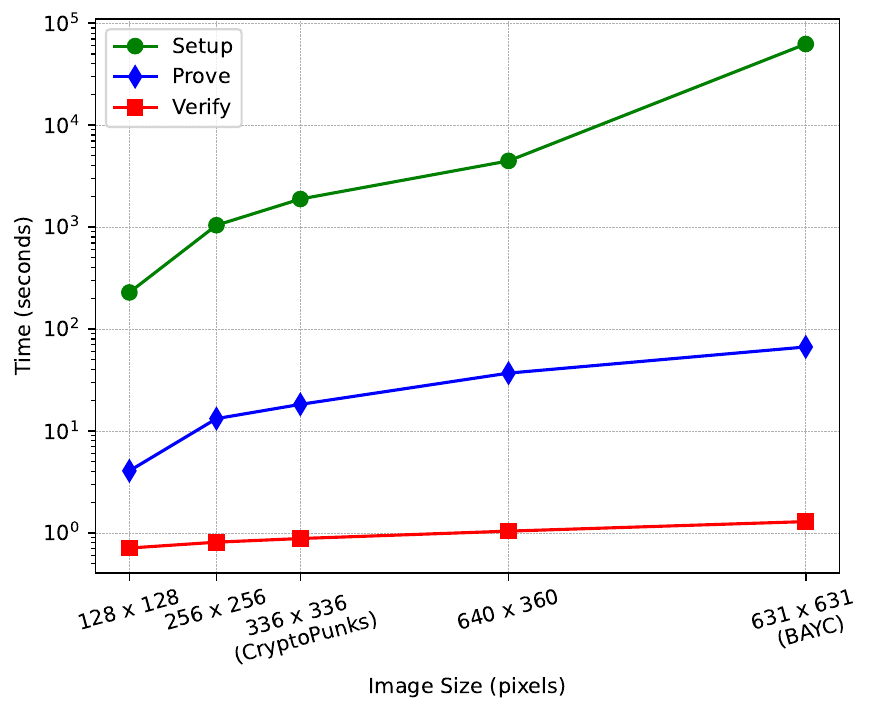}
        \caption{Time consumption (in logarithmic scale) during the proving and verification phases of \zksnark generation in Alg.~\ref{alg:adv} line 3 using Groth16 and during the Setup phase (i.e., $\zksnarkscheme.\zksnarksetup$ of Groth16), for images of varying sizes.}
        \label{fig:Time}
    \end{subfigure}
    \hfill
    \begin{subfigure}[b]{0.49\linewidth}
        \centering
        \includegraphics[width=\linewidth]{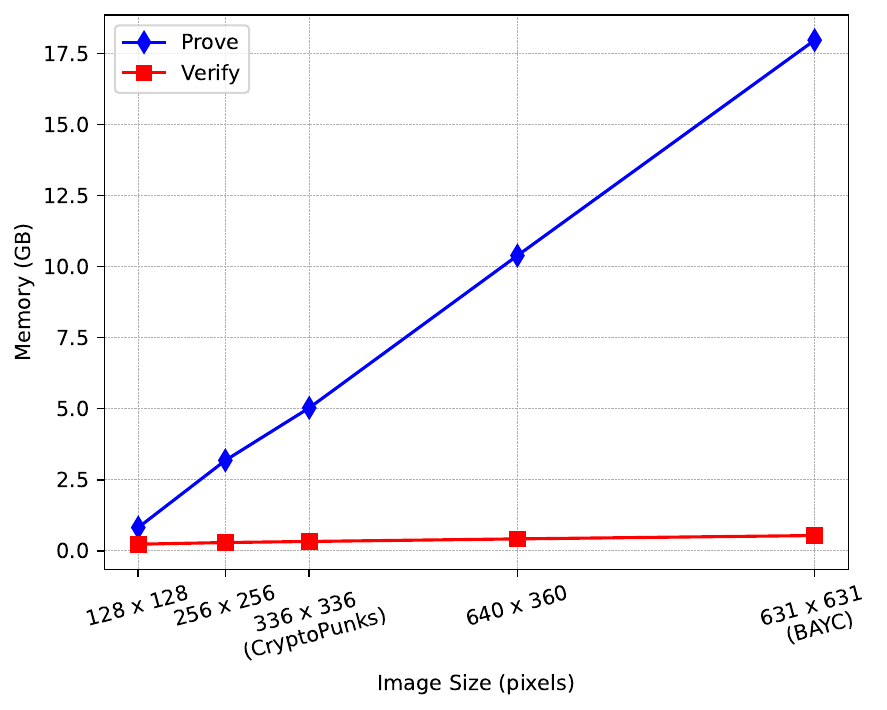}
        \caption{Memory consumption (in logarithmic scale) during the proving and verification phases of \zksnark generation in Alg.~\ref{alg:adv} line 4 using Groth16, for images of varying sizes.}
        \label{fig:Memory}
    \end{subfigure}
    \caption{Resource consumption during \zksnark generation and verification.}
\end{figure}

\paragraph{Protocol setup and proof generation.} 
Fig.~\ref{fig:Time} shows the time necessary to perform the setup, generation, and verification of the \zksnark. The memory usage is reported in Fig.~\ref{fig:Memory}. The setup phase is largely the most time-consuming and memory-intensive, however, this operation is performed only once
for a given image size and resize factor, its output
does not contain any private information and can be thus
publicly shared. Therefore, the setup phase can also be
computed on the cloud without violating the confidentiality of the private image. In Fig.~\ref{fig:Memory}, we did not include setup memory because it constantly saturates all available RAM, relying on swap memory when necessary.
 The proof generation is the second most demanding operation. It is performed only once for a given preview, and it must be computed locally by the seller to preserve confidentiality. Remarkably, the verification of the proof that is performed by each potential buyer, is the most efficient.

\CR{
\paragraph{Avoiding  demanding setup.}
As described, the setup phase relies on the generation of a trusted CRS and commitment key, which in our experiments took approximately one day to compute on standard hardware. This cost is amortized by allowing the same setup to be reused for future exchanges of the same asset type. However, the setup time may still present a barrier for practical deployment, especially in applications requiring frequent updates or new asset types.

One way to mitigate this cost, as we said before, is to consider alternative zk-SNARK constructions that can be instantiated in the RO model. In such approaches, the CRS is replaced using cryptographic hash functions modeled as random oracles, removing the need for a trusted party and potentially reducing setup time. Recent works, such as~\cite{EC:BCRSVW19,EC:CHMMVW20,EC:ChiOjhSpo20}, demonstrate practical efficient zk-SNARKs in the RO model, which could be adapted to our protocol.
Another direction is to explore universal or updatable SNARKs, where a single setup supports many statements or can be efficiently updated for new relations~\cite{EPRINT:GabWilCio19}. These constructions can further reduce the setup overhead when protocol parameters or asset types change, if the RO is not an option.
While for the commitment key one can rely on standard tecniques such as ``hashing into the curve''~\cite{TCHES:WahBon19} in case of e.g. Pedersen commitment  (and model the hash function as a RO) to  avoid the presence of a trusted third party.
}
\CR{\paragraph{Dealing with large images.}
A valid criticism of our feasibility results concerns the size of the images used in our experiments, which may not reflect the high-resolution images typically exchanged in centralized services (e.g., on Getty Images, it is possible to buy and sell images with resolutions up to $\sim$20 MP). The main bottleneck preventing us from conducting experiments with high-resolution images was the need to generate the common reference string for Groth16 (the Setup in Fig.~\ref{fig:Time}) with only $32$ GB of RAM, which required utilizing swap memory, significantly increasing execution time.

While  our experiments are focused on smaller image sizes, it is important to consider that in the literature, there exists a rich line of work that addresses the problem of generating {\zksnark}s for high-resolution images, on statements comparable to those in our experiments. These works explore ad-hoc techniques (e.g., exploiting the internal structure of images, namely the RGB channels) for generating such proofs even with limited resources, enabling the computation of a \zksnark for images of almost any size. Specifically, recent works~\cite{SP:DCB25,EPRINT:DEH24} demonstrate that it is possible to compute these proofs for images $\sim$10 MP within two hours, with a peak memory usage of approximately 10 GB.

Lastly, in~\cite{SP:DVVZ25}, the authors demonstrated that by relaxing the confidentiality requirements of the image, it is possible to design an efficient proof system for images, requiring only 4 GB of RAM to compute a proof for images up to $\sim$30 MP, in the order of minutes.
That is, by directly leveraging~\cite{SP:DVVZ25}, our approach becames inherently scalable in large-scale markets: each user can independently generate proofs tailored to their specific requirements and computational resources, enabling the protocol to accommodate a broad class of users, including those with limited hardware. Users can flexibly produce proofs for images of practically any size, with computation time proportional to the resources available to them.
}


\paragraph{The smart contract.}
We developed a smart contract in Solidity that extends the ERC-721 standard and implements the logic of the fair exchange as described in Fig.~\ref{fig:exprot} and Alg.~\ref{alg:complaint}.
The key exchange, along with key pair generation, has been implemented using the secp256k1 curve. We use SHA-3  as a heuristic for the random oracle.
In our experiments, we evaluate the cost  of running the main functions of the proposed smart contract that implements $\Courfeprot$ as described in Sec.~\ref{sec:constructions} both on Ethereum and on a layer 2 technology. The following functions correspond to the phases detailed in Sec.~\ref{sec:constructions}. Below, we briefly describe the purpose of each function.
The \texttt{Deployment} function refers to the cost of deploying the smart contract.
\texttt{Mint} creates a new NFT with a pointer (to an IPFS directory) to the corresponding \zksnark (along with its statement). Additionally, the smart contract stores in the state the commitment of the symmetric key needed to decrypt the asset (i.e., $\commassetkey$), resembling the operation at line 4 of Alg.~\ref{alg:adv}.
The \texttt{Bid} function is invoked by the buyer to make an offer, where the buyer also publishes their (ephemeral) public key on the smart contract. This operation resembles lines 1 to 3 of the fair exchange detailed in Fig.~\ref{fig:exprot}.
The \texttt{Accept} function is invoked by the seller to accept the offer and deliver the encryption key, following the steps from lines 4 to 10 in Fig.~\ref{fig:exprot}.
Finally, we developed two functions that allow the buyer to conclude the exchange by either accepting the value sent by the seller or raising a complaint to the smart contract. Both functions correspond to lines 11 to 13 of Fig.~\ref{fig:exprot}. The buyer can invoke the \texttt{Confirm} function to successfully conclude the exchange (this function allows both parties to conclude the exchange quickly, avoiding unnecessary delays) or the \texttt{Complain} function, which implements the logic described in Alg.~\ref{alg:complaint}.
The results in terms of fees are summarized in Fig.~\ref{fig:SCexp}.
\begin{figure}[!htb]
\centering
\resizebox{.73\linewidth}{!}{%
\begin{tabular}{l|lrr}
\multicolumn{1}{c|}{\textbf{Function}} & \textbf{Gas} & \textbf{Ethereum fees (in \$)} & \textbf{Layer 2 fees (in \$)} \\ \hline
\texttt{Deployment} & 2818661 & 117.54  & 0.00021 \\ \hline
\texttt{Mint}       & 269247  & 11.23   & 0.00002 \\ \hline
\texttt{Bid}        & 161394  & 6.73    & 0.000012 \\ \hline
\texttt{Accept}     & 331532  & 13.83   & 0.000024 \\ \hline
\texttt{Confirm}    & 69240   & 2.89    & 0.0000051 \\ \hline
\texttt{Complain}   & 1633536 & 68.12   & 0.00012
\end{tabular}%
}
\caption{The costs associated with the main functions of the proposed smart contract extending the ERC-721 standard. The Layer 2 solution we selected for comparison is Optimism. The gas price was taken from October 10, 2024. }
\label{fig:SCexp}
\end{figure}
Note that, as expected, if the fair exchange is successfully carried on, the costs are significantly lower than in the case of complaints.
We highlight that, according to Fig.~\ref{fig:SCexp} the decision to deploy the contract on Ethereum mainnet or on a Layer 2 technology can be influenced by the value of the asset for which the NFT is created, in relation to the management costs. The deployment of high-value assets may justify the costs of Ethereum mainnet, while lower-value assets could benefit from the cost of Layer 2 solutions. 

\CR{
\paragraph{Broader applicability beyond NFT images.}
While our experimental evaluation focuses on NFTs representing images, the underlying protocol is readily adaptable to a wide spectrum of digital assets. According to our experiments, what we effectively demonstrate and benchmark is the ability to generate privacy-preserving proofs for data represented as a multidimenional matrix, which is de facto the standard format for images and can be naturally extended to other asset types (e.g., audio, PDF files) or in general other multidimensional data, where the representation follows a similar structure.}
\section*{Acknowledgments}
This work was partially supported by
project SERICS (PE00000014) under the MUR National Recovery and Resilience Plan funded by the European Union - NextGenerationEU.
Ivan Visconti is member of the Gruppo Nazionale Calcolo Scientifico-Istituto Nazionale di Alta Matematica (GNCS-INdAM) and his research contribution on his work was in part financially supported under the National Recovery and Resilience Plan (NRRP), Mission 4, Component 2, Investment 1.1, Call for tender No. 104 published on 2.2.2022 by the Italian Ministry of University and Research (MUR), funded by the European Union -NextGenerationEU - Project Title ``PARTHENON'' - CUP D53D23008610006 - Grant Assignment Decree No. 959 adopted on June 30, 2023 by the Italian Ministry of Ministry University and Research (MUR).

\bibliographystyle{splncs04}
\bibliography{bibliography/biblio,bibliography/abbrev3,bibliography/crypto}

\begin{thebibliography}{10}
\providecommand{\url}[1]{\texttt{#1}}
\providecommand{\urlprefix}{URL }
\providecommand{\doi}[1]{https://doi.org/#1}

\bibitem{FCW:ADMM14}
Andrychowicz, M., Dziembowski, S., Malinowski, D., Mazurek, L.: Fair two-party computations via bitcoin deposits. In: B{\"o}hme, R., Brenner, M., Moore, T., Smith, M. (eds.) FC 2014 Workshops. {LNCS}, vol.~8438, pp. 105--121. Springer, Berlin, Heidelberg (Mar 2014)

\bibitem{EC:AsoShoWai98}
Asokan, N., Shoup, V., Waidner, M.: Optimistic fair exchange of digital signatures (extended abstract). In: Nyberg, K. (ed.) EUROCRYPT'98. {LNCS}, vol.~1403, pp. 591--606. Springer, Berlin, Heidelberg (May~/~Jun 1998)

\bibitem{cryptoeprint:2025/388}
Baecker, R., Gerhart, P., Katz, J., Schröder, D.: Fair exchange for decentralized autonomous organizations via threshold adaptor signatures. Cryptology {ePrint} Archive, Paper 2025/388 (2025)

\bibitem{ESORICS:BanDziMal16}
Banasik, W., Dziembowski, S., Malinowski, D.: Efficient zero-knowledge contingent payments in cryptocurrencies without scripts. In: Askoxylakis, I.G., Ioannidis, S., Katsikas, S.K., Meadows, C.A. (eds.) ESORICS~2016, Part~II. {LNCS}, vol.~9879, pp. 261--280. Springer, Cham (Sep 2016)

\bibitem{bauer2022erc}
Bauer, D.P.: Erc-721 nonfungible tokens. In: Getting Started with Ethereum : A Step-by-Step Guide to Becoming a Blockchain Developer, pp. 55--74. Apress, Berkeley, CA (2022)

\bibitem{EC:BCRSVW19}
{Ben-Sasson}, E., Chiesa, A., Riabzev, M., Spooner, N., Virza, M., Ward, N.P.: Aurora: Transparent succinct arguments for {R1CS}. In: Ishai, Y., Rijmen, V. (eds.) EUROCRYPT~2019, Part~I. {LNCS}, vol. 11476, pp. 103--128. Springer, Cham (May 2019)

\bibitem{C:BenKum14}
Bentov, I., Kumaresan, R.: How to use bitcoin to design fair protocols. In: Garay, J.A., Gennaro, R. (eds.) CRYPTO~2014, Part~II. {LNCS}, vol.~8617, pp. 421--439. Springer, Berlin, Heidelberg (Aug 2014)

\bibitem{HTLC}
Bitcoin, W.: {Hash Time Locked Contracts} (2024)

\bibitem{CCS:CGGN17}
Campanelli, M., Gennaro, R., Goldfeder, S., Nizzardo, L.: Zero-knowledge contingent payments revisited: Attacks and payments for services. In: Thuraisingham, B.M., Evans, D., Malkin, T., Xu, D. (eds.) ACM CCS 2017. pp. 229--243. {ACM} Press (Oct~/~Nov 2017)

\bibitem{EC:CHMMVW20}
Chiesa, A., Hu, Y., Maller, M., Mishra, P., Vesely, P., Ward, N.P.: Marlin: Preprocessing {zkSNARKs} with universal and updatable {SRS}. In: Canteaut, A., Ishai, Y. (eds.) EUROCRYPT~2020, Part~I. {LNCS}, vol. 12105, pp. 738--768. Springer, Cham (May 2020)

\bibitem{EC:ChiOjhSpo20}
Chiesa, A., Ojha, D., Spooner, N.: Fractal: Post-quantum and transparent recursive proofs from holography. In: Canteaut, A., Ishai, Y. (eds.) EUROCRYPT~2020, Part~I. {LNCS}, vol. 12105, pp. 769--793. Springer, Cham (May 2020)

\bibitem{STOC:Cleve86}
Cleve, R.: Limits on the security of coin flips when half the processors are faulty (extended abstract). In: 18th ACM STOC. pp. 364--369. {ACM} Press (May 1986)

\bibitem{SP:DCB25}
Datta, T., Chen, B., Boneh, D.: { VerITAS: Verifying Image Transformations at Scale }. In: 2025 IEEE Symposium on Security and Privacy (SP). pp. 97--97. IEEE Computer Society, Los Alamitos, CA, USA (May 2025)

\bibitem{SP:DVVZ25}
{Della Monica}, P., Visconti, I., Vitaletti, A., Zecchini, M.: { Trust Nobody: Privacy-Preserving Proofs for Edited Photos with Your Laptop }. In: 2025 IEEE Symposium on Security and Privacy (SP). pp. 14--14. IEEE Computer Society, Los Alamitos, CA, USA (May 2025)

\bibitem{EC:DGGK21}
Dobraunig, C., Grassi, L., Guinet, A., Kuijsters, D.: Ciminion: Symmetric encryption based on {Toffoli}-gates over large finite fields. In: Canteaut, A., Standaert, F.X. (eds.) EUROCRYPT~2021, Part~II. {LNCS}, vol. 12697, pp. 3--34. Springer, Cham (Oct 2021)

\bibitem{EPRINT:DEH24}
Dziembowski, S., Ebrahimi, S., Hassanizadeh, P.: {VIMz}: Verifiable image manipulation using folding-based {zkSNARKs}. Cryptology ePrint Archive, Paper 2024/1063 (2024), accepted at Privacy Enhancing Technologies Symposium (PETS) 2025.

\bibitem{CCS:DziEckFau18}
Dziembowski, S., Eckey, L., Faust, S.: {FairSwap}: How to fairly exchange digital goods. In: Lie, D., Mannan, M., Backes, M., Wang, X. (eds.) ACM CCS 2018. pp. 967--984. {ACM} Press (Oct 2018)

\bibitem{ASIACCS:EckFauSch20}
Eckey, L., Faust, S., Schlosser, B.: {OptiSwap}: Fast optimistic fair exchange. In: Sun, H.M., Shieh, S.P., Gu, G., Ateniese, G. (eds.) ASIACCS 20. pp. 543--557. {ACM} Press (Oct 2020)

\bibitem{EC:FucWol24}
Fuchsbauer, G., Wolf, M.: Concurrently secure blind schnorr signatures. In: Joye, M., Leander, G. (eds.) EUROCRYPT~2024, Part~II. {LNCS}, vol. 14652, pp. 124--160. Springer, Cham (May 2024)

\bibitem{EPRINT:GabWilCio19}
Gabizon, A., Williamson, Z.J., Ciobotaru, O.: {PLONK}: Permutations over {Lagrange}-bases for oecumenical noninteractive arguments of knowledge. Cryptology ePrint Archive, Report 2019/953 (2019)

\bibitem{EC:GarKiaLeo15}
Garay, J.A., Kiayias, A., Leonardos, N.: The bitcoin backbone protocol: Analysis and applications. In: Oswald, E., Fischlin, M. (eds.) EUROCRYPT~2015, Part~II. {LNCS}, vol.~9057, pp. 281--310. Springer, Berlin, Heidelberg (Apr 2015)

\bibitem{CCS:GMMMTT22}
Glaeser, N., Maffei, M., Malavolta, G., {Moreno-Sanchez}, P., Tairi, E., Thyagarajan, S.A.K.: Foundations of coin mixing services. In: Yin, H., Stavrou, A., Cremers, C., Shi, E. (eds.) ACM CCS 2022. pp. 1259--1273. {ACM} Press (Nov 2022)

\bibitem{USENIX:GKRRS21}
Grassi, L., Khovratovich, D., Rechberger, C., Roy, A., Schofnegger, M.: Poseidon: {A} new hash function for zero-knowledge proof systems. In: Bailey, M., Greenstadt, R. (eds.) USENIX Security 2021. pp. 519--535. {USENIX} Association (Aug 2021)

\bibitem{EC:Groth16}
Groth, J.: On the size of pairing-based non-interactive arguments. In: Fischlin, M., Coron, J.S. (eds.) EUROCRYPT~2016, Part~II. {LNCS}, vol.~9666, pp. 305--326. Springer, Berlin, Heidelberg (May 2016)

\bibitem{cryptoeprint:2025/059}
Hafezi, H., Partap, A., Das, S., Bonneau, J.: Fair signature exchange. Cryptology {ePrint} Archive, Paper 2025/059 (2025)

\bibitem{ESORICS:HLTWW21}
He, S., Lu, Y., Tang, Q., Wang, G., Wu, C.Q.: Fair peer-to-peer content delivery via blockchain. In: Bertino, E., Shulman, H., Waidner, M. (eds.) ESORICS~2021, Part~I. {LNCS}, vol. 12972, pp. 348--369. Springer, Cham (Oct 2021)

\bibitem{FileBounty}
Janin, S., Qin, K., Mamageishvili, A., Gervais, A.: Filebounty: Fair data exchange. In: 2020 IEEE European Symposium on Security and Privacy Workshops (EuroS\&P W). IEEE (Sep 2020)

\bibitem{TCC:KMTZ13}
Katz, J., Maurer, U., Tackmann, B., Zikas, V.: Universally composable synchronous computation. In: Sahai, A. (ed.) TCC~2013. {LNCS}, vol.~7785, pp. 477--498. Springer, Berlin, Heidelberg (Mar 2013)

\bibitem{RSA:KupLys10}
K{\"u}p{\c c}{\"u}, A., Lysyanskaya, A.: Usable optimistic fair exchange. In: Pieprzyk, J. (ed.) CT-RSA~2010. {LNCS}, vol.~5985, pp. 252--267. Springer, Berlin, Heidelberg (Mar 2010)

\bibitem{CCS:LYHMGZ21}
Li, Y., Ye, C., Hu, Y., Morpheus, I., Guo, Y., Zhang, C., Zhang, Y., Sun, Z., Lu, Y., Wang, H.: {ZKCPlus}: Optimized fair-exchange protocol supporting practical and flexible data exchange. In: Vigna, G., Shi, E. (eds.) ACM CCS 2021. pp. 3002--3021. {ACM} Press (Nov 2021)

\bibitem{smartZKCP}
Liu, X., Zhang, J., Wang, Y., Yang, X., Yang, X.: {SmartZKCP}: Towards practical data exchange marketplace against active attacks. Cryptology ePrint Archive, Paper 2024/941 (2024), accepted at International Conference on Blockchain Research and Applications (BCRA) 2024.

\bibitem{Maxwell15}
Maxwell, G.: Zero knowledge contingent payment (2015)

\bibitem{SP:TaiMorMaf21}
Tairi, E., {Moreno-Sanchez}, P., Maffei, M.: {A$^2$L}: Anonymous atomic locks for scalability in payment channel hubs. In: 2021 {IEEE} Symposium on Security and Privacy. pp. 1834--1851. {IEEE} Computer Society Press (May 2021)

\bibitem{CCS:TSZMKB24}
Tas, E.N., Seres, I.A., Zhang, Y., Melczer, M., Kelkar, M., Bonneau, J., Nikolaenko, V.: Atomic and fair data exchange via blockchain. In: Luo, B., Liao, X., Xu, J., Kirda, E., Lie, D. (eds.) ACM CCS 2024. pp. 3227--3241. {ACM} Press (Oct 2024)

\bibitem{SP:ThyMal21}
Thyagarajan, S.A.K., Malavolta, G.: Lockable signatures for blockchains: Scriptless scripts for all signatures. In: 2021 {IEEE} Symposium on Security and Privacy. pp. 937--954. {IEEE} Computer Society Press (May 2021)

\bibitem{SP:ThyMalMor22}
Thyagarajan, S.A.K., Malavolta, G., {Moreno-Sanchez}, P.: Universal atomic swaps: Secure exchange of coins across all blockchains. In: 2022 {IEEE} Symposium on Security and Privacy. pp. 1299--1316. {IEEE} Computer Society Press (May 2022)

\bibitem{CCS:VanSonThy24}
Vanjani, N., Soni, P., Thyagarajan, S.A.K.: Functional adaptor signatures: Beyond all-or-nothing blockchain-based payments. In: Luo, B., Liao, X., Xu, J., Kirda, E., Lie, D. (eds.) ACM CCS 2024. pp. 1493--1507. {ACM} Press (Oct 2024)

\bibitem{TCHES:WahBon19}
Wahby, R.S., Boneh, D.: Fast and simple constant-time hashing to the {BLS12-381} elliptic curve. {IACR} {TCHES}  \textbf{2019}(4),  154--179 (2019)

\end{thebibliography}

\appendix
\renewcommand\theHsection{Appendix \thesection}
\newpage
\section{Preliminaries}\label{app:preliminaries}
\subsection{Notation}
\label{sec:notation}

We denote by $\secp \in \mathbb{N}$ the security parameter.
We denote by $\boolset^n$ the set of all $n$-bit long strings and by $\boolset^*$ the set of all binary strings.
Given two bit strings $a$ and $b$ we denote with $a\|b$ a new bit string that is the concatenation of the string $a$ and the string $b$.
Given a message space $\mathcal{M} $, we denote by $|\mathcal{M}| \in \N$  the bit-length of the representation of the elements in the message space $\mathcal{M}$.

If $S$ is a finite set, we denote by $x \leftarrow S$ the process of sampling $x$ from  $S$, and by 
$x \randomfromset S$ a random and uniform one. 
A function $\negl$
is negligible if it vanishes faster than the inverse of
any polynomial (i.e.,  for any constant $c > 0$ 
for sufficiently large $\secp$ it holds that
$\negl(\secp) \leq {\secp^{-c}}$). 
Throughout, $\G$ denotes an Abelian group of prime order $q$, for notational convenience we use additive notation for the binary operations in $\G$.

We say that two (set of) distributions $X = \{X (\secp)\}_{\lambda \in \N}$ and
$Y = \{Y (\secp)\}_{\lambda \in \N}$  are \emph{computationally indistinguishable}  
if every probabilistic polynomial-time ($\ppt$) algorithm $\dist$
cannot distinguish $X$ and $Y$ except with negligible probability. 
We denote it as $X \cind Y$.

When we explicitly specify the random tape $\rho$ for a randomized algorithm $\mathsf{A}$, then
we write $a' \leftarrow \mathsf{A}(a; \rho)$ to indicate that $\mathsf{A}$ outputs $a'$ given input $a$ and random tape $\rho$.
Given a scheme $\Pi_{\mathsf{A}}$ composed by a sequence of algorithms  $(\mathsf{Alg}_1,\ldots,\mathsf{Alg}_n)$, we will sometimes refer to a specific  algorithm with $\Pi_{\mathsf{A}}.\mathsf{Alg}_j$ for some $1 \leq j\leq n$. 

In our constructions, we will use a function $\RO{\ell}:\boolset^*\rightarrow\boolset^\ell$ that is concretely instantiated through a cryptographic hash function, but that is heuristically and ideally considered a random oracle (i.e., a completely random function) in the analysis.

\subsection{Cryptographic Primitives}

\paragraph{Symmetric Encryption scheme.} 
A symmetric encryption scheme enables parties to securely communicate by encrypting and decrypting messages using a shared secret key. More formally:
\begin{definition}
 A symmetric encryption scheme for a key space $\mathcal{K}_e$, a plaintext space $\mathcal{M}_e$, and a ciphertext space $\mathcal{C}_e$  is a tuple of algorithms $\encscheme=(\enckeygen,\allowbreak\encenc,\allowbreak\encdec)$.
 \begin{itemize}
     \item $k \leftarrow \enckeygen(1^\secp)$ is a randomized algorithm that takes as input the security parameter $\secp$, and it outputs a secret key $k \in \mathcal{K}_e$.
     \item $c \leftarrow \encenc(k,m)$  is an  algorithm that takes as input the secret key $k \in \mathcal{K}_e$ and a plaintext $m \in\mathcal{M}_e$, and it outputs a ciphertext $c \in\mathcal{C}_e$.
     \item $m \leftarrow \encdec(k,c)$ is a deterministic algorithm that takes as input the secret key $k \in\mathcal{K}_e$ and a plaintext $c \in\mathcal{C}_e$, and it outputs a plaintext $m \in\mathcal{M}_e$.
 \end{itemize}
A symmetric encryption scheme need to satisfy the following properties:
\oursubparagraph{Correctness.} For any message $m \in\mathcal{M}_e$:
\[
\PRC{
  k \leftarrow \enckeygen(1^\secp)\\
  c \leftarrow \encenc(k,m)\\
}{
  m = \encdec(k,c)
}=1
\]
\oursubparagraph{Semantic security.} For every $\ppt$ adversary $\adv$ there exists a negligible function such that:
\[
\left|
\frac{1}{2} -
\PRC{
 k \leftarrow \enckeygen(1^\secp)\\
 (m_0,m_1) \leftarrow \adv(1^\secp)\\
 b\randomfromset\boolset; c_b \leftarrow \encenc(k,m_b)\\
 b' \leftarrow \adv(c_b)
}{
 (m_0,m_1) \in \mathcal{M}_e \land \\
 b' = b
}\right|\leq \negl(\secp)
\]
\end{definition}

\paragraph{Commitment schemes.}
\label{sec:commscheme}
A commitment scheme allows a sender to create a commitment to a secret value. It may later open the commitment and reveal the value or some information about the value in a verifiable manner. More formally:

\begin{definition}
A (non-interactive) commitment scheme is a tuple of algorithms $\commscheme = (\commkeygen,\commcommit, \commopen)$.

    \begin{itemize}
        \item $\commkey \leftarrow \commkeygen(1^\secp)$ is a deterministic algorithm that takes as input the security parameter $\secp$, and it outputs a committment public key $\commkey$. This $\commkey$ specifies the message space  $\mathcal{M}_c$, a commitment space $\mathcal{C}_c$ and a randomness space $\mathcal{R}_c$. 

        \item $c \leftarrow \commcommit(\commkey,m)$ is a randomized algorithm that takes as input the committment public key $\commkey$ and a message $m$, and it outputs a commitment $c$ for the message. 

        When we write $c \leftarrow \commcommit(\commkey,m; r)$ with $r \in \mathcal{R}_c$ we fix the random tape of the algorithm, and thus with the same $m$ and $r$, we always obtain the same sequence of bits for $c$.

        \item $\boolset \leftarrow \commopen(\commkey,c,m,r)$ is a deterministic algorithm that takes as input the committment public key $\commkey$, a commitment $c$, a message $m$ and a randomness $r$, and it outputs $1$ when accepting the commitment, otherwise outputs $0$, rejecting it.

    \end{itemize}
A commitment scheme satisfies the following properties:

\oursubparagraph{Correctness.} For any message $m \in \mathcal{M}_c$:
\[
\PRC{ \commkey \leftarrow \commkeygen(1^\secp)\\ 
      c \leftarrow \commcommit(\commkey,m)}
    {\commopen(\commkey,c,m,r) = 1 } = 1
\]

\oursubparagraph{Binding.} For every $\ppt$ adversary $\adv$ there exists a negligible function such that:
\[
\PRC{\commkey \leftarrow \commkeygen(1^\secp)\\
     (c,m,m',r,r')\leftarrow\adv(\commkey)\\
    }
    {m \neq m' \\\land 
     \commopen(\commkey,c,m,r) = 1  \\\land
     \commopen(\commkey,c,m',r') = 1
    }
    \leq \negl(\secpar)
\]

\oursubparagraph{Hiding.} For any $m,m' \in \mathcal{M}_c$ and for every $\ppt$ adversary $\adv$:
\[
    \left|\PRC{
        \commkey \leftarrow \commkeygen(1^\secp)\\ c \leftarrow\commcommit(\commkey,m)
    }{
    \adv(c) = 1
    }
    -
    \PRC{
        \commkey \leftarrow \commkeygen(1^\secp)\\ c' \leftarrow\commcommit(\commkey,m')
    }{
    \adv(c') = 1
    }\right| \leq \negl(\secp)
\]
\end{definition}

Throughout the paper, we use an hash-based commitment scheme, thus $\commcommit(\commkey,m;r) \defeq \RO{\kappa}(m\|r)$ such that $\kappa = |\mathcal{C}_c|$\footnote{\label{ftn:ROinCircuit}
For efficiency, we use this commitment scheme that is known to be secure in the random oracle model. We will compute a  \zksnark proving the correct computation of a commitment through this scheme. Thus, to avoid the use of \zksnark for proving a claim over a hash function used to instantiate a random oracle, we require that the instantiated hash-based commitment scheme is secure as it is (i.e., without random oracles). A similar heuristic assumption is for instance used in~\cite{EC:FucWol24}.}.

\paragraph{Signature schemes.}\label{sec:sign}
A signature scheme allows a signer to produce a digital signature for a message such that everyone can check the authenticity of the message and be sure that the signature came from only that specific signer. More formally:

\begin{definition}
A signature scheme is a tuple of algorithms $\signscheme = (\signkeygen,\signsign,\allowbreak\signverify)$.

\begin{itemize}
    \item $(\sk,\pk) \leftarrow \signkeygen(1^\secp)$ is a randomized algorithm that takes as input the security parameter $\secp$, and it outputs a signature key pair $(\sk,\pk)$. The $\pk$ includes the message space $\mathcal{M}_s$ and the signature space $\mathcal{S}_s$.

    \item $\sigma \leftarrow \signsign(\sk,m)$ is a randomized algorithm that takes as input the signing key $\sk$ and the message $m \in \mathcal{M}_s$, and it outputs the signature $\sigma \in \mathcal{S}_s$ for the message.

    \item $\boolset \leftarrow \signverify(\pk,m,\sigma)$ is a deterministic algorithm that takes as input the $\pk$, a message $m \in \mathcal{M}_s$ and a signature $\sigma \in \mathcal{S}_s$, and it outputs $1$,  if the verification is successful, otherwise it outputs $0$.
\end{itemize}
A signature scheme must satisfy the following properties:

\oursubparagraph{Correctness.} For any message $m \in \mathcal{M}_s$:
\[
\PRC{(\sk,\pk) \leftarrow \signkeygen(1^\secp)\\\sigma \leftarrow \signsign(\sk,m)}{\signverify(\pk,m,\sigma) = 1} = 1
\]

\oursubparagraph{Unforgeability (EUF-CMA).} For every $\ppt$ adversary $\adv$ there exists a negligible function such that:
\[
\PRC{(\sk, \pk) \leftarrow \signkeygen(1^\secp) \\
     (m,\sigma) \leftarrow \adv^{\oracle_\signsign(\sk,\cdot)}(\pk)
    }
    {m \not \in Q \land\\
     \signverify(\pk,m,\sigma) = 1
    }
    \leq \negl(\secp)
\]
where $\oracle_\signsign(\sk,\cdot)$ is a signature oracle with which $\adv$ can interact and obtains signatures for messages of its choice, and  $Q$ is the set of all messages that are given as input to the signature oracle.
\end{definition}

\paragraph{Zero-Knowledge Succint Non-Interactive Argument of Knowledge.}
Let a polynomial relation $\rel$ be a function $\rel: \boolset^* \times \boolset^* \rightarrow \boolset$. We call $\stat$ the statement and $\wit$ the witness for the relation $\rel$, and given the tuple $(\stat,\wit)$ allows one to efficiently verify whether $\rel(\stat,\wit) = 1$.

\begin{definition}
    A Zero-Knowledge Succinct Non-Interactive Argument of Knowledge (\zksnark), for a polynomial relation $\rel$ is a tuple of algorithms $\zksnarkscheme = (\zksnarksetup, \zksnarkprove, \zksnarkverify)$.

    \begin{itemize}
        \item $\crs \leftarrow \zksnarksetup(1^\secp)$ is a randomized algorithm that takes as input the security parameter $\secp$, and it outputs a common reference string $\crs$.

        \item $\pi \leftarrow \zksnarkprove(\crs, \stat, \wit)$ is a randomized algorithm that takes as input the  common reference string $\crs$, an instance $\stat$ of the statement, and a witness $\wit$, such that $\rel(\stat,\wit) = 1$. It outputs a proof $\pi$.

        \item $\boolset \leftarrow \zksnarkverify(\crs,\stat,\pi)$ is a deterministic algorithm that takes the  common reference string $\crs$, an instance $\stat$ and a proof $\pi$, and it outputs $1$ if the verification is successful, otherwise it outputs $0$.
    \end{itemize}
    A ZK-snark need to satisfy the following properties:

    \oursubparagraph{Completeness.} For any pair $(\stat,\wit)$ such that $\rel(\stat,\wit) = 1$:
    \[
    \PRC{\crs \leftarrow \zksnarksetup(1^\secp)\\
         \pi \leftarrow \zksnarkprove(\crs,\stat,\wit)
        }
        {\zksnarkverify(\crs,\stat,\pi) = 1}
    = 1
    \]

    \oursubparagraph{Succinctness.} The proof size is $\poly(\secp + \log(|\wit|))$ and the verifier running time is $\poly(\secp + \log(|\wit|) + |\stat|)$.

    \oursubparagraph{Knowledge Soundness.} Considering an auxiliary input distribution $\mathcal{Z}$, if for every $\ppt$ adversary $\adv$ there exists a $\ppt$ extractor $\Ext$ such that:
    \[
    \PRC{\crs  \leftarrow \zksnarksetup(1^\secp)\\
         \aux_\mathcal{Z} \leftarrow \mathcal{Z}(\crs)\\
         (\stat,\pi) \leftarrow \adv(\crs,\aux_\mathcal{Z})\\
         \wit \leftarrow \Ext(\crs,\aux_\mathcal{Z})
        }
        {\zksnarkverify(\crs,\stat,\pi) = 1 \land \\
         \rel(\stat,\wit) = 0
        }
        \leq \negl(\secp)
    \]

    \oursubparagraph{(Composable) Zero Knowledge.} If there exists a simulator $\Sim = (\Sim_{kg}, \Sim_{prv})$ such that the following conditions hold for all $\ppt$ adversaries $\adv$:
    \begin{itemize}
        \item {Keys Indistinguishability.}
        \begin{align*}
        &\left|
        \PRC
        {\crs  \leftarrow \zksnarksetup(1^\secp)}
        {\adv(\crs) = 1}\right.
        \\
        &\left. -\PRC
        {(\crs, \trapdoor_{kg}) \leftarrow \Sim_{kg}(1^\secp)}
        {\adv(\crs) = 1}
        \right| \leq \negl(\secp)
        \end{align*}

        \item {Proof Indistinguishability.} For all $(\stat,\wit)$ such that $\rel(\stat,\wit) = 1$:
        \begin{align*}
        &\left|\PRC
        {(\crs, \trapdoor_{kg}) \leftarrow \Sim_{kg}(1^\secp)}
        {\pi \leftarrow \zksnarkprove(\crs,\stat,\wit),\\
         \adv(\crs,\pi) = 1
        }\right.
        \\
        &\left. -\PRC
        {(\crs, \trapdoor_{kg}) \leftarrow \Sim_{kg}(1^\secp)}
        {\pi \leftarrow \Sim_{prv}(\crs,\stat,\trapdoor_{kg}),\\
         \adv(\crs,\pi) = 1
        }\right| \leq \negl(\secp)
        \end{align*}
    \end{itemize}
\end{definition}

\subsection{Fair Exchange}\label{sec:fair_ex}
In this section, we recall the definition of fair exchange that comes from previous works~\cite{EC:AsoShoWai98,RSA:KupLys10}, instantiating it for our use case.

We have two parties, \emph{Alice} and \emph{Bob}, who want to exchange their digital assets $i_A$ and $i_B$. The traditional way to solve this problem is to rely on an \emph{Arbiter}, a trusted third party (TTP), which is assumed to be honest and will help Alice and Bob exchange the assets fairly. 
For the sake of simplicity, we assume that the parties are capable of directly verifying that the digital asset corresponds to what they actually desire, without the need to introduce any other actors, (as instead it is the case of~\cite{EC:AsoShoWai98,RSA:KupLys10}). 

Before the fair exchange protocol is executed, we assume that each party generates a pair of keys of $\signscheme$  generated with the $\signscheme.\signkeygen$ algorithm and publishes the public keys so that all parties can verify the authenticity of the communications. 

\begin{definition}
A fair exchange protocol $\feprot$ is a three-party communication protocol where: Alice and Bob run two algorithms $\mathsf{A}$ and $\mathsf{B}$ to exchange digital assets, and the Arbiter runs another trusted algorithm $\mathsf{T}$ to regulate the exchange. Alice runs on inputs $i_A$ and a verification algorithm $V_A$ and Bob runs on inputs $i_B$ and a verification algorithm $V_B$.
Namely,
    $[\alpha, \beta] \leftarrow [\mathsf{A}(i_A, V_A), \mathsf{B}(i_B, V_B), \mathsf{T}].$

A fair exchange protocol must satisfy the following properties:
 \oursubparagraph{Completeness.}   the execution of the protocol by honest parties results in Alice giving in output Bob’s digital asset and vice versa. More formally:
    $\operatorname{Pr} [[i_B, i_A] \leftarrow [\mathsf{A}(i_A, V_A),\allowbreak \mathsf{B}(i_B, V_B), \mathsf{T}]] = 1.$ 
 \oursubparagraph{Fairness.} A fair exchange protocol is \emph{fair} if at the end of the protocol, either both Alice and Bob get the digital assets of the other party, or in case a honest party among Alice and Bob does not give in output the asset of the other party, then the other party does not get any non-trivial information\footnote{Informally, if a party misbehaves so that the other party does not receive the asset, then the misbehaving party will learn nothing about the other party's asset, apart from the information that was known before the start of the protocol.} about the asset of the other party.     
    More formally, an exchange protocol is \emph{fair} if for all $\ppt$ malicious $\mathsf{B}^*$ there exists a simulator $\Sim_\mathsf{B^*}$ with oracle access to $\mathsf{T}$ such that:
    $\{
       \beta \leftarrow[\Sim_{\mathsf{B}^*}^\mathsf{T}(i_B,V_B)] : \beta
    \}\cind
    \{
      [\bot,\beta] \leftarrow [\mathsf{A}(i_A, V_A), \mathsf{B}^*(i_B, V_B), \mathsf{T}] :\beta 
    \}$
    
  
    A dual condition must hold for any $\ppt$ malicious $\mathsf{A}^*$.
\end{definition}
$V_A$ (respectively, $V_B$) is the verification algorithm run by Alice (respectively, Bob) to evaluate whether $i_B$ (respectively, $i_A$) is the desired asset. For instance, if $i_B$ is supposed to be an artwork created by a known artist, $V_A$ verifies, among other tasks, the artist's signature on the artwork. 

The definition of fair exchange in~\cite{EC:AsoShoWai98,RSA:KupLys10} also involves two other trusted parties: a \emph{Tracker} and a \emph{Bank}. The former is used to make sure that the assets exchanged by the parties are the correct ones, while the latter takes care of eventual payments and money exchanges. The verification of the digital assets is executed through the Tracker in an off-line phase, where parties receive from the Tracker their verification algorithms. For simplicity, we deviate from previous work by lightening the definition; we focus exclusively on verification algorithms that allow the parties to autonomously verify that digital assets match their expectations, and we assume that $\arbiter$ also takes the role of a \emph{Bank}.

\paragraph{Fair Exchange for digital assets on a blockchain.}
Assume that Alice is a seller of a certain digital asset (i.e., $i_A$), and Bob is a buyer who will exchange money (i.e., $i_B$) for Alice's asset.
If $i_A$ and $i_B$ are implemented via blockchain-based digital assets, then one can leverage the assumption that the blockchain is a trusted ``entity'' and use it as the arbiter $\arbiter$ in a fair exchange protocol. Indeed, it accepts messages from $\mathsf{A}$ and $\mathsf{B}$ and executes its instructions honestly (i.e., according to the logic of the smart contract implementing the arbiter).
These types of fair exchange over a blockchain are defined using \emph{smart contracts}.

\section{Proof of Theorem~\ref{thm:opti-constr}}\label{app:theorem-proof} The correctness of the protocol can be verified by inspection.

\oursubparagraph{Seller Fairness.}
According  to the seller fairness property of Def.~\ref{def:ourfe}, a malicious buyer $\buyer^*$ does not learn anything about $\priv$ when $\seller$ aborts. In $\Courfeprot$, $\seller$ aborts when $V_\seller$ outputs $0$ (i.e., when it receives an offer that does not satisfy its request). In this case, the only public information available for $\buyer^*$ is $\pub$ and, therefore, we want to show that, from $\pub$, $\buyer^*$ does not learn anything about $\priv$. According to Def.~\ref{def:ourfe}, we must show that everything that $\buyer^*$ learns about $\priv$ from $\pub$ can be simulated. $Sim_{\buyer^*}^\arbiter$ is realized as follows. 
Given as input 
    the  asset preview $ f(\asset^\star)$ and
    $\token_{\mathsf{thr}} \in \N^+$,
    $\Sim_{\buyer^*}^\arbiter$ does the following:
    
    \begin{enumerate*}[label=(\textbf{\arabic*})]
        \item $\Sim_{\buyer^*}^\arbiter$ executes $\Sim_{kg}(1^\secp)$ and it obtains the pair $(\crs,\trapdoor_{kg})$;
        \item $\Sim_{\buyer^*}^\arbiter$ executes $\commscheme.\commkeygen(1^\secp)$ and it obtains $\commkey$;
        \item $\Sim_{\buyer^*}^\arbiter$ computes the pair $(\trustparam \defeq (\crs,\allowbreak\commkey),\allowbreak \trapdoor \defeq \trapdoor_{kg})$;
        \item $\Sim_{\buyer^*}^\arbiter$ executes $\encscheme.\enckeygen(1^\secp)$ and it obtains $k$;
        \item $\Sim_{\buyer^*}^\arbiter$ computes a ``garbage'' value $g \randomfromset \boolset^{|\mathcal{M}_e|}$
        (note that $\Sim_{\buyer^*}^\arbiter$ knows $\mathcal{M}_e$, because $\encscheme$ is defined for a given message space);
        \item $\Sim_{\buyer^*}^\arbiter$ executes  $\encscheme.\encenc(k,g)$ and it obtains $e_g$;
        \item $\Sim_{\buyer^*}^\arbiter$ executes  $\commscheme.\commcommit(\commkey,k)$ and it obtains $c_k$;
        \item $\Sim_{\buyer^*}^\arbiter$ sets $x = (e_g,\assetprev,c_k,\commkey)$, then it executes $\Sim_{prv}(\crs,x,\trapdoor_{kg})$ and it obtains $\pi$;
        \item $\Sim_{\buyer^*}^\arbiter$ executes $\advsubroutine(\token_{\mathsf{thr}})$ and it obtains $V_\seller$ and $V_\buyer$;
        \item $\Sim_{\buyer^*}^\arbiter$ sets $\pub = (x,\pi,V_\seller, V_\buyer)$ and executes $\store(\pub)$;
        \item $\Sim_{\buyer^*}^\arbiter$ executes $\buyer^*(\tokens{n},\trustparam,\pub)$ and outputs whatever $\buyer^*$ outputs. 
        
        Note that, from Def.~\ref{def:ourfe}, $\buyer^*$ runs with $\tokens{n}$ as input, where $\tokens{n}$ come from the wallet $\wallet_\buyer$ of $\buyer^*$. $\Sim_{\buyer^*}^\arbiter$ obtains $\wallet_\buyer$ simply by querying $\arbiter$. Furthermore, $\Sim_{\buyer^*}^\arbiter$ runs in \emph{polynomial time}, since it executes $\buyer^*$ that is a polynomial time algorithm and executes other operations that can be performed in polynomial time. 
    \end{enumerate*}

Now, in this proof, we refer to a real game experiment $\mathsf{RG}$ as the game played by a $\ppt$ adversary $\buyer^*$ in which it tries to generate an output $\beta$ that is distinguishable from the output produced by the simulator  $\Sim_{\buyer^*}^\arbiter$. As required by the definition, we want to show that the probability that, in $\mathsf{RG}$, $\buyer^*$ produces a $\beta$ distinguishable from the output of $\Sim_{\buyer^*}^\arbiter$ is negligible. The proof proceeds by contradiction,  (i.e., there exists a $\ppt$ adversary $\buyer^*$ winning in $\mathsf{RG}$ with probability at least $\lambda^{-c}$, for some constant $c$) and uses the following hybrid experiments.

The first hybrid is $\mathcal{H}_1$ and it corresponds to $\mathsf{RG}$ except for, in  $\Courfeprot.\ourfesetup$ algorithm, the generation of $\crs$ composing the trusted parameters $\trustparam$ that is computed by $\Sim_{kg}(1^\secp)$ instead of $\zksnarkscheme.\zksnarksetup$. The success probability of $\buyer^*$ in $\mathcal{H}_1$ is only negligibly far from the one in $\mathsf{RG}$; otherwise, there is an obvious reduction that breaks the key indistinguishability of the ZK in $\zksnarkscheme$. 


The second hybrid is $\mathcal{H}_2$ where the experiment proceeds as in $\mathcal{H}_1$ except that, in $\Courfeprot.\ourfeadv$, instead of running $\encscheme.\encenc$ on $\asset || \assetsign$, $\encscheme.\encenc$ is run on a ``garbage'' value $g \randomfromset \boolset^{|\mathcal{M}_e|}$. The success probability of $\buyer^*$ in $\mathcal{H}_2$ is only negligibly far from the one in $\mathcal{H}_1$; otherwise, there is an obvious reduction that breaks the semantic security of $\encscheme$. 

The third hybrid is $\mathcal{H}_3$ where the experiment proceeds as in $\mathcal{H}_2$ except that, in $\Courfeprot.\ourfeadv$,  $\commscheme.\commcommit$ is computed over a value different from $\assetkey$. The success probability of $\buyer^*$ in $\mathcal{H}_3$ is only negligibly far from the one in $\mathcal{H}_2$; otherwise, there is an obvious reduction that breaks the hiding of $\commscheme$.

The fourth hybrid is $\mathcal{H}_4$ where the experiment proceeds as in $\mathcal{H}_3$ except for, in $\Courfeprot.\ourfeadv$, the generation of the proof $\pi$ that is computed by $\Sim_{prv}(\crs,x,\allowbreak\trapdoor_{kg})$ instead of $\zksnarkscheme.\zksnarkprove$. The success probability of $\buyer^*$ in $\mathcal{H}_4$ is only negligibly far from the one in $\mathcal{H}_3$; otherwise, there is an obvious reduction that breaks the proof indistinguishability of the ZK in $\zksnarkscheme$. 


Note that the experiment $\mathcal{H}_4$ executes the same operations made by $\Sim_{\buyer^*}^\arbiter$ from step (\textbf{1}) to step (\textbf{8}). Subsequent steps (\textbf{9}) and (\textbf{10}) are deterministic operations, while in step (\textbf{11}), $\Sim_{\buyer^*}^\arbiter$ outputs whatever $\buyer^*$ outputs. Hence, the success probability of $\buyer^*$ in $\mathcal{H}_4$ is the same success probability that $\Sim_{\buyer^*}^\arbiter$ has to produce a $\beta$ distinguishable from another execution of the same algorithm $\Sim_{\buyer^*}^\arbiter$. This is clearly not possible and, thus, since $\mathcal{H}_4$ and $\mathsf{RG}$ are indistinguishable, it contradicts through hybrid arguments that $\buyer^*$ can win the game with non-negligible probability. 

\oursubparagraph{Buyer Fairness.}
According to the buyer fairness property (see Def.~\ref{def:ourfe}), we consider a protocol instance of $\Courfeprot$ where there is an honest buyer $\buyer$, with  wallet $\wallet_\buyer$, such that $\tokens{n} = \{\token_i\}_{1 \leq i \leq n} $ and $ \tokens{n} \intk \wallet_\buyer$. In the following proof, we show that for any $\ppt$ malicious seller $\seller^*$,
if $\buyer$ aborts the fair exchange, then  the probability that there is  a $j \in \{1,\ldots,n\}$ such that $\token_j \not \in \wallet_\buyer$ is negligible.

The proof goes by contradiction. Suppose that there exists a $\ppt$ malicious seller $\seller^*$ of $\Courfeprot$ such that the following holds:
\[
    \PRC
    {
        \trustparam \leftarrow \ourfesetup(1^\lambda)\\
        (\priv,\pub) \leftarrow \seller^*(\trustparam)\\
        
        [a, \bot] \leftarrow [\seller^*(\priv), \buyer(\tokens{n}), \arbiter](\trustparam,\pub)
    }
    {
        \exists i \in \{1, \ldots, n\}\text{ s.t. } \token_i \not \in \wallet_\buyer
    }\geq \secp^{-c}
\]
for infinitely many values of $\secp$ (i.e., $\seller^*$ is successful) where $c$ is a non-negative constant. In the following, we will show that the above non-negligible probability allows us to reach a contradiction.

According to $\Courfeprot$ (see Fig.~\ref{fig:exprot} and Alg.~\ref{alg:complaint}) there are two events (that happens one after another) where $\buyer$ has the possibility to abort. Both events occur after $\buyer$ executes the following operation $\tx(\seller^*,\tokens{n},t')$, namely $\buyer$ locks $\tokens{n}$ into $\arbiter$ until $t'$ and after $t'$ the tokens are
transferred to $\seller^*$.

We analyze such events in the following:
\begin{enumerate}
    \item $\buyer$ aborts after line 2 of Alg.~\ref{alg:complaint}. This event happens if the timestamp $t_\mathsf{complain}$\footnoteref{ftn:tcomp} (loosely speaking a timeout for $\seller^*$) is passed and $\encassetkey $ is equal to $ \bot$ (i.e., after the timeout $\seller^*$ did not send $\encassetkey$).

    Since both the time  (see Sec.~\ref{sec:blockchain}) and the value of $\encassetkey$ are managed  by $\arbiter$, the event that $t > t_{\mathsf{comp}}$ and $\encassetkey$ is equal to $\bot$, but $\arbiter$ does not give back the $\tokens{n}$ to $\buyer$,
    causing the buyer $\buyer$ to abort without $\tokens{n} \intk \wallet_\buyer$ happens only if $\arbiter$ does not behave correctly, leading us to a contradiction, because we have assumed that $\arbiter$ behaves correctly (i.e., according to the behavior described in Sec.~\ref{sec:constructionNFT}).

    \item We are left with the case where $\buyer$ aborts after line 10 of Alg.~\ref{alg:complaint}. Since $\buyer$ behaves correctly and $\arbiter$ executes its operation correctly (in accordance with  the previous point) the event that $\arbiter$ does not give back $\tokens{n}$ to $\buyer$,
    causing the buyer $\buyer$ to abort without $\tokens{n} \intk \wallet_\buyer$ happens only because, for a message $m \in \mathcal{M}_e$ and a key $k \in \mathcal{K}_e$, $\encscheme.\encdec(k, \encscheme.\encenc(k,m)) \neq m$, but this contradicts the correctness of $\encscheme$,
    or because, during the computation of the commitment $\seller^*$ computes  $m,m' \in \mathcal{M}_c$,  $r,r' \in \mathcal{R}_c$, and $c \in \mathcal{C}_c$ such that $m \neq m'$ and $\commscheme.\commopen(\commkey,c,m,r) = \commscheme.\commopen(\commkey,c,m',r') = 1$ but this contradicts the binding of $\commscheme$.
    
\end{enumerate}
Note that in $\Courfeprot$, the buyer $\buyer$ can also abort before starting the exchange with the seller, specifically after evaluating $\pub$. This event occurs even before $\buyer$ locks $\tokens{n}$ in $\buyer$ and, in this scenario, it would trivially contradict the correctness of $\arbiter$, as $\arbiter$ would arbitrarily remove $\tokens{n}$ from $\wallet_\buyer$. Furthermore, $\buyer$ could potentially evaluate  $\pub$ without having any tokens in $\wallet_\buyer$.

\oursubparagraph{Broadcast Communication only.}
We argue that all communication between the seller and the buyer is carried out exclusively via broadcast messages towards~$\arbiter$, that is implemented through a decentralized platform. In particular, every message exchanged in the protocol, including commitments, encrypted assets, Diffie-Hellman messages, ciphertexts, and dispute-related data, is either posted directly to~$\arbiter$ or stored in a publicly accessible medium with a corresponding pointer submitted to~$\arbiter$. No direct or private communication between the parties is required at any stage of the protocol. Note that, no other party observing the messages exchanged by the seller and the buyer can obtain $(\assetkey\|r)$ otherwise it is able to break the Decisional DH assumption. This property is immediate from the construction and can be verified by inspecting the sequence of messages posted to~$\arbiter$ and the referenced public data.

This concludes the proof of this theorem.
\hfill$\square$

\section{Considerations on \texorpdfstring{$\ourfeprot$}{the definition}}\label{app:notes-on-definition}

\paragraph{Definition for a symmetric exchange.}
A natural question that arises when considering Def.~\ref{def:ourfe} is whether it is possible to easily extend our definition to the ``symmetric case'', where both parties are required to advertise and then exchange an asset. The answer is affirmative, and the generalized definition can be easily derived by looking at Def.~\ref{def:ourfe}. Specifically, in the ``symmetric case'', the advertisement phase (i.e., the $\ourfeprot.\ourfeadv$ algorithm) is executed by both parties, disclosing their public information obtained from $\ourfeprot.\ourfeadv$ and then they engage in the three-party execution, exchanging their private information defined by executions of $\ourfeprot.\ourfeadv$.

With respect to the fairness property, it becomes even simpler. In fact, it is sufficient to ensure that the current Seller Fairness property holds for both parties.
\section{Considerations on \texorpdfstring{$\Courfeprot$}{the construction}}\label{app:notes-on-contruction}

\paragraph{On the information maintained by $\arbiter$.}
In the construction $\Courfeprot$, for simplicity, we have so far assumed that $\arbiter$ has the ability to store all $\pub$. However, there are cases in our construction (i.e., when $\arbiter$ is a blockchain) where this may not be feasible, or where doing so would require the smart contract,  associated with $\arbiter$ to be prohibitively expensive. Specifically, publishing $\pub$  could result in significant costs in terms of fees. A clear example of this arises from the fact that $\pub$ contains an instance of the statement of the \zksnark generated during the advertisement in Alg.~\ref{alg:adv}. Among other values, this instance includes $\assetenc$, the encryption of the asset $\asset^\star$. Naturally, the size of $\assetenc$ increases with the size of $\asset^\star$, and in contexts where $\asset^\star$ is the order of MBs (e.g., $\asset^\star$ is a digital painting), it becomes impractical to store this information within a smart contract.

However, upon further inspection of the information required by $\arbiter$ to ensure a fair protocol (see Alg.~\ref{alg:complaint}), we highlight that $\arbiter$ does not need to know all the details contained in $\pub$. Rather, it only needs access to $\commkey, \commassetkey$, which is the information related to the commitment to the encryption key $\assetkey$. This information is small. That is, we allow $\arbiter$ storing only $\commkey, \commassetkey$ while storing a pointer to the remaining information in $\pub$. Specifically, the additional information could be stored externally (e.g., on IPFS
), with $\arbiter$ pointing to it.

\paragraph{On the optimistic nature of \texorpdfstring{$\Courfeprot$}{our construction}.}
The construction $\Courfeprot$ can be considered as an optimistic construction. Specifically, recalling that the advertisement phase is unique and publicly shared among all potential buyers, and focusing our attention on the fair exchange outlined in Fig.~\ref{fig:exprot}, we observe that if the subroutine invoked by the buyer $\buyer$ (i.e., $V_\buyer$) outputs the value $1$, thus indicating that the key $\assetkey$ is correct (i.e., $\commscheme.\commopen(\commkey,\commassetkey,\assetkey, r) \neq 1$), the honest buyer has no reason to complain to $\arbiter$, thus concluding the fair exchange. 

However, analyzing Fig.~\ref{fig:exprot}, we note that $\arbiter$ is still invoked by both parties three additional times, even if the buyer does not complain about receiving an incorrect $\assetkey$. Specifically, the buyer $\buyer$ interacts with $\arbiter$ in lines 2 and 3, and the seller $\seller$ in line 7. Analyzing these operations, we see that both $\buyer$ and $\seller$ deposit their public key in line 2 and in line 7, respectively, using $\arbiter$ essentially as a broadcast channel to provide this information to the other party. Regarding the operation of locking funds in line 3, we observe that the buyer locks the payment into $\arbiter$ for a time $t'$, requiring (almost) no computation from $\arbiter$. 

Thus, although $\arbiter$ is involved in the exchange process even if both parties behave correctly, we can still consider the proposed construction as optimistic. This is because the work of $\arbiter$ remains minimal: $\arbiter$ does not need to perform any active computation unless the buyer $\buyer$ files a complaint about the malicious behavior of the seller. We anticipate that the experiments in Sec.~\ref{sec:experiments} confirmed this, as the results clearly demonstrate that the costs (in terms of gas) associated with the management of the NFT are entirely dominated by the buyer's complaint operation (see Fig.~\ref{fig:SCexp} for a quick anticipation of the experimental results).

\paragraph{Considering old asset owners.}
The construction $\Courfeprot$ is designed to be compatible with the management of confidential and authenticated NFTs. In this application context, when a user sells an NFT, she transfers the ownership of such an NFT along with the possibility for the new owner to fully enjoy the asset $\asset$ linked to the NFT. According to the adopted standards for NFT trading (e.g., ERC-721), only the current owner of an NFT obtains the tokens for a sale. This means that past owners have no interest in maintaining confidential the encryption key $\assetenc$ used to encrypt $\asset$. 

We can mitigate this problem by leveraging a reasonable game-theoretic assumption and assume that participants behave rationally. For instance, we can extend standard smart contracts with an incentive mechanism for old owners, allowing them to earn a portion of the sale price of all the subsequent purchases. In this way, we discourage them to disclose the encryption key because if the asset $\asset$ starts circulating over other users, the demand and, therefore, the sale price for new purchase would drastically decrease. 

Note that, this issues is verified for the case of NFT due to the concept of property, however, for other application contexts our construction is perfectly aligned with the real-world scenario (e.g., in centralized marketplace like Getty Images is possible to buy a photo and, then, share it with others).

\section{Considerations on experiments}\label{app:notes-on-experiments}

\paragraph{Main technical choices.}
The proof  described in line 4 of Alg.\ref{alg:adv} has been computed with (1) Circom to write the circuits, (2) Snarkjs to set up the proof and  (3) Rapidsnark to generate the proof leveraging Groth16~\cite{EC:Groth16} as the underlying proof system. Groth16  is commonly used in blockchain applications due to its succinctness,  providing constant proof size and verification time. 
The signature scheme is ECDSA with Poseidon 128~\cite{USENIX:GKRRS21}. In our experiment, we compute only the hash of the image because the overhead of verifying the signature on the (hashed) image is constant and introduces negligible time and memory consumption compared to the other operations performed in the circuit. Poseidon 128 is also used for the commitment of the encryption key\footnoteref{ftn:ROinCircuit}.
For the symmetric encryption of the image, we use Ciminion~\cite{EC:DGGK21}, specifically relying on its reference implementation in Circom\footnote{\url{https://github.com/kudelskisecurity/circom-ciminion}}.
To evaluate the \zksnark circuit, we encoded the input according to the approach proposed by Khovratovich\footnote{\href{https://hackmd.io/@7dpNYqjKQGeYC7wMlPxHtQ/BkfS78Y9L}{https://hackmd.io/@7dpNYqjKQGeYC7wMlPxHtQ/BkfS78Y9L}}. We rely on the Circom implementation of Poseidon 128, with the number of field elements of a single Poseidon Sponge iteration set to 16, which is the maximum available value.
For all other parameters, we refer the reader to the Circom official repository\footnote{\href{https://github.com/iden3/circomlib/blob/master/circuits/poseidon.circom}{https://github.com/iden3/circomlib/master/circuits/poseidon.circom}}.

\end{document}